\documentclass{JHEP3}
 \usepackage{epsfig}
 \usepackage{graphicx}
 \usepackage{enumerate}
 \usepackage{enumerate}
 \usepackage{verbatim}
 
 \def\be{\begin{equation}}
 \def\bea{\begin{eqnarray}}
 \def\ee{\end{equation}}
 \def\eea{\end{eqnarray}}
 \newcommand{\beq}{\begin{equation}}
 \newcommand{\eeq}{\end{equation}}
 \newcommand{\beqa}{\begin{eqnarray}}
 \newcommand{\eeqa}{\end{eqnarray}}
 \newcommand{\beqar}{\begin{eqnarray*}}
 \newcommand{\eeqar}{\end{eqnarray*}}

 \def\openone{\leavevmode\hbox{\small1\kern-3.3pt\normalsize1}}
 \def\roughly#1{\raise.3ex\hbox{$#1$\kern-.75em\lower1ex\hbox{$\sim$}}}

\title{Collisions with Black Holes and Deconfined Plasmas}

\author{Aaron J. Amsel, Donald Marolf, and Amitabh Virmani \\
Department of Physics, University of California at Santa Barbara,
Santa Barbara, CA 93106, USA \\ E-mail:
 \it{amsel@physics.ucsb.edu, marolf@physics.ucsb.edu, virmani@physics.ucsb.edu} }

\abstract{We use AdS/CFT to investigate {\it i}) high energy
collisions with balls of deconfined plasma surrounded by a confining
phase and {\it ii}) the rapid localized heating of a deconfined
plasma. Both of these processes are dual to collisions with black
holes, where they result in the nucleation of a new ``arm'' of the
horizon reaching out in the direction of the incident object. We
study the resulting non-equilibrium dynamics in a universal limit of
the gravitational physics which may indicate universal behavior of
deconfined plasmas at large $N_c$. Process ({\it i}) produces
``virtual'' arms of the plasma ball, while process ({\it ii}) can
nucleate surprisingly large bubbles of a higher temperature phase.}


\begin{document}

\section{Introduction}

Studies of gravity/gauge duality \cite{Juan,MAGOO, wit} have given
insight into many aspects of strongly coupled gauge theories. A
central aspect of this correspondence is that deconfined plasmas in
the gauge theory are dual to black holes on the gravity side  of
the correspondence \cite{wit}. Although no gravity dual is known
for gauge theories of direct experimental interest (e.g., QCD), the
universal properties of black holes in gravity suggest that insights
obtained from gravity/gauge duality may describe universal aspects
of deconfined plasmas at large
 $\lambda$, $N_c$, where $N_c$ is the number of colors and $\lambda$ is the 't Hooft coupling.
The known utility of the large $N_c$ expansion in QCD then further
suggests that such dualities may help to explain the physics
observed at heavy ion colliders \cite{shear_viscosity,
bulk_viscosity, jet, far}. See e.g., \cite{MatRev} for a recent
review.

Much of the literature on this subject has focused on using black
holes to explore equilibrium thermodynamic properties of deconfined
plasmas (such as entropies, free energies, and phase transitions,
see e.g., \cite{wit, equilibrium}) or quasi-static near-equilibrium
properties such as hydrodynamic transport coefficients, see e.g.,
\cite{shear_viscosity, bulk_viscosity, jet}.  While transport
coefficients can describe important hydrodynamics (such as elliptic
flow, see e.g., \cite{Shuryak}), such processes must be quasi-static
in the sense that each local region of the plasma remains close to
thermal equilibrium.

However, one may also ask about more dynamic processes.  One example
is the formation of a plasma from collisions studied in \cite{far}.
We will be interested here in related (but different) settings,
involving the rapid transfer of energy to {\it pre-existing}
plasmas. At low rates of energy transfer, these processes result in
a flow of heat described  by hydrodynamic thermal conductivity,
perhaps associated with some expansion or hydrodynamic flow of the
plasma. But what happens at higher rates of energy transfer?

The result is most easily seen from the gravity dual. Recall that
the most interesting feature of the dual black hole is its event
horizon and that this horizon is a surface defined by null
geodesics.   A small energy flux across the black hole horizon leads
only to local expansion of the horizon in a manner readily described
by hydrodynamics (see e.g., \cite{membrane}).  On the other hand,
larger fluxes lead to more dramatic results.  As an extreme case,
consider the collision of a small black hole with a larger black
hole as shown in Fig.~\ref{podfig}.
 %\begin{figure}
 %\begin{center}
 %\includegraphics[width=1.5 in]{arm.eps}
 %\end{center}
 \FIGURE{\epsfig{file=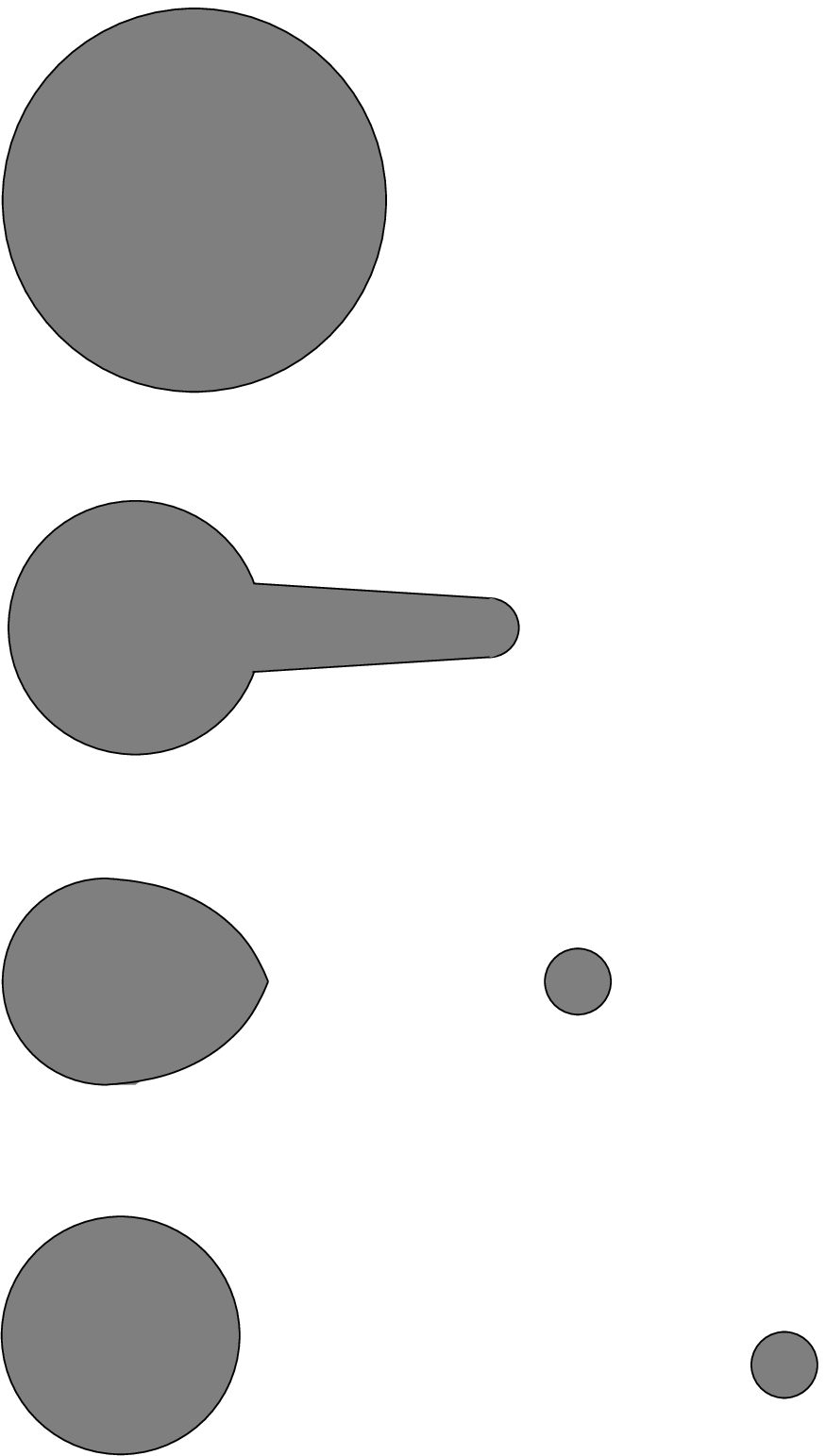, width=1.5 in}
 \caption{This cartoon depicts a time sequence of the creation of
an ``arm" as a small object falls into a large black hole.  Note
that the arm forms {\it before} the arrival of the incident object.}
 \label{podfig} }
 %\end{figure}

At first, the two black holes approach each other, each maintaining
its own event horizon with only minor effects from the other black
hole. However, when the black holes are sufficiently close together
a new common horizon suddenly forms around both black holes,
resulting in a single highly deformed black hole.  In particular,
the region between the two black holes is now enclosed in a new
``arm'' extending from the large black hole.  The arm is then
absorbed by the black hole as the system relaxes to equilibrium.

This picture is currently being investigated in numerical
simulations of black hole collisions \cite{US}, though the analogous
phenomenon for collisions of equal mass black holes is already well
illustrated by numerical simulations, see e.g.,
\cite{EarlyNumerics}. In any case, it has long been known on general
grounds (see e.g., \cite{HE,Waldbook}) that the new arm of the
horizon forms along a spacelike caustic; i.e., the new arm of the
black hole effectively nucleates ``from scratch." Furthermore, the
incident object need not be another black hole for this to occur;
any sufficiently rapid flux of incoming energy results in the
formation of a similar black hole arm that reaches out toward the
incoming object. As discussed in \cite{AMV} (which generalizes the
arguments of \cite{membrane}), in any spacetime dimension $d$ the
threshold for such nucleation to occur is\footnote{This result holds
for a homogeneous object much smaller than the curvature scale at
the black hole horizon whose cross-section transverse to the motion
is round.  The size of the object along the direction of motion does
not affect the result so long as the collision takes place in less
than the relaxation time $1/T$ of the black hole.}
\begin{equation}
\frac{E}{A}  \sim  \frac{T}{G_d} \, ,
\end{equation}
where $E,A$ are the energy and transverse area of the incident
object, $T$ is the temperature of the target black hole, $G_d$ is
Newton's gravitational constant, and we have set $\hbar =1$.

The purpose of this paper is to calculate the shape of the
nucleating arm and to interpret the results in terms of deconfined
plasmas; we shall be less concerned with the final relaxation to
equilibrium. As we discuss in more detail below, the formation of a
black hole arm can correspond to either of two processes for the
plasma, depending on the direction in which the arm extends. If the
black hole arm reaches out along one of the gauge theory directions,
then our process is naturally interpreted as the formation of a
``virtual" arm of deconfined plasma catalyzed by the incident flux
of energy. If, on the other hand, the black hole arm extends in the
holographic direction (which corresponds to energy scales in the
gauge theory), then it represents a strong local heating of the
plasma via, for example, the decay of an excited quasi-particle of
large mass\footnote{One can see from the gravity dual that  such
quasi-particles exist at large $N_c, \lambda$ even in the deconfined
phased.  These quasi-particles correspond to bulk gravitons,
strings, or localized black holes raised sufficiently far above the
horizon of the black hole associated with the deconfinement
transition.}.
 Such heating takes the plasma far from local equilibrium and
is not well-described by hydrodynamics.  In particular, when the
system allows further phase transitions, local heating can create a
bubble of a higher temperature phase within the original plasma.

Though the details of specific models are certain to contain much
interesting physics, our goal here is to take only the first steps
toward understanding this dynamics in the context of gauge/gravity
duality.  For this reason, we focus here on universal or
near-universal properties of collisions with black holes (and thus
presumably with deconfined plasmas). In particular, we use the fact
that the geometry sufficiently close to the horizon of any black
hole of non-zero temperature can be approximated by Rindler space,
which is just flat Minkowski space in accelerated coordinates.   We
also take the incident object to be both small and highly boosted so
that horizon dynamics can be studied through the simple exercise of
tracing null geodesics through the Aichelburg-Sexl metric \cite{AS}.

This analysis is performed in section \ref{geodesics}. We then
interpret the results in terms of large $N_c$ gauge theories in
sections \ref{pbs} and \ref{mesons}. Section \ref{pbs} considers
high energy collisions with plasma balls and the resulting virtual
arms that extend in the gauge theory directions. Section
\ref{mesons} considers the case of rapid local heating.   As a
particular example we examine certain gauge theories with flavor in
which stable mesons can exist within the original plasma. In that
case, our local heating can cause the mesons to melt within a
localized bubble, though the bubbles turn out to be surprisingly
large.  Gross properties of such bubbles are studied at both strong
and weak 't Hooft couplings.  We conclude with a brief discussion in
section \ref{disc}.

 \section{How far can a black hole reach?}
 \label{geodesics}

 In this section we address universal
properties of collisions with black holes.  Our primary goal is to
characterize the black hole arm that nucleates to engulf an incident
object.  We wish to determine the shape of this arm, and in
particular its thickness and length.  Section \ref{setup} discusses
the physical setup and determines the resulting horizon, while
section \ref{extract} extracts the desired parameters.

 \subsection{Physical and mathematical framework}
 \label{setup}

Our study of collisions with black holes is carried out under two
key approximations, within which the behavior is universal. First,
we use the fact that the geometry sufficiently close to the horizon
of any black hole with temperature $T \neq 0$ can be approximated by
Rindler space, which is just flat Minkowski space in accelerated
coordinates. Second, we consider the limit where the object
approaches the black hole at high velocity so that generic
(uncharged, non-spinning) objects are well-described by the simple
Aichelburg-Sexl metric \cite{AS}. Studying horizon dynamics during
the collision then reduces to tracking null geodesics (the so-called
generators of the event horizon) as they propagate through the
associated spacetime.  Such geodesics were studied in \cite{FPV},
which we review below for completeness.

%\begin{figure}%\begin{center}%\includegraphics[width=3.5in]{ztplane.eps}%\end{center}
 \FIGURE{\epsfig{file=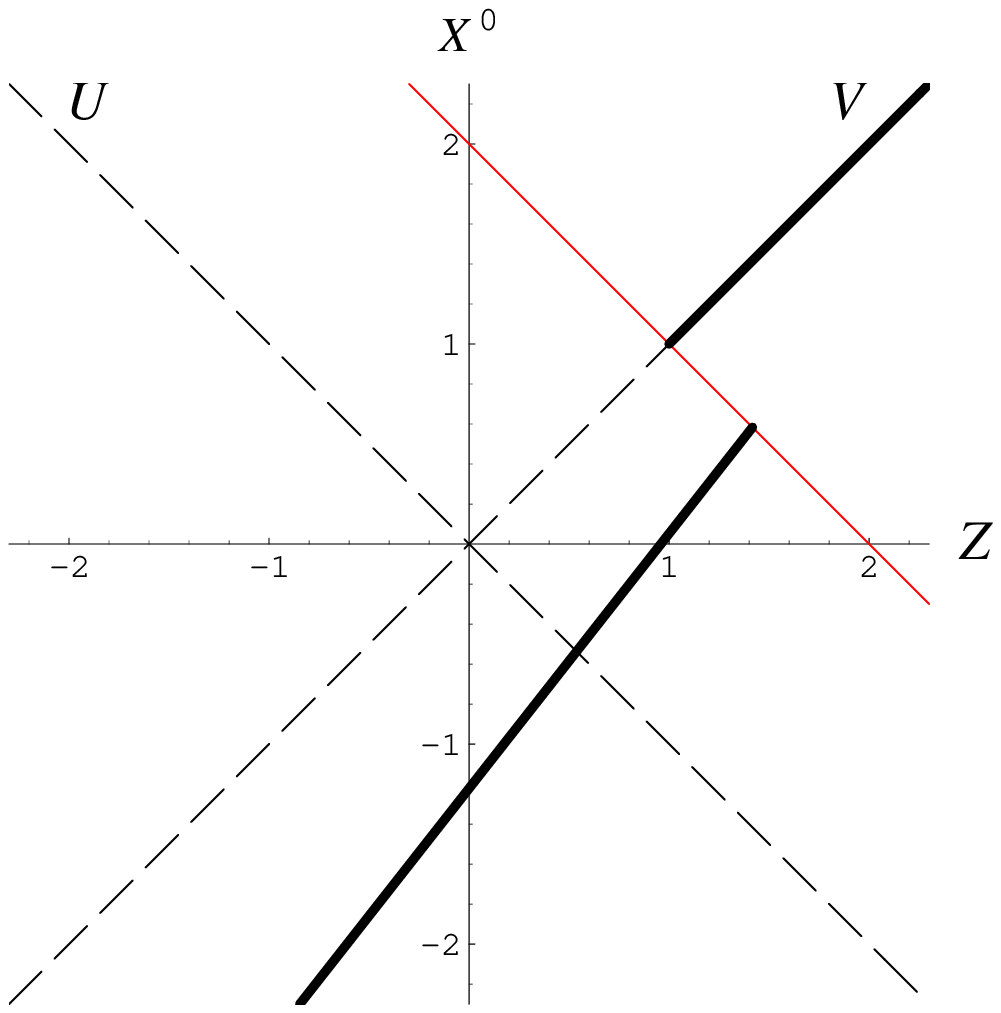, width=2.5in} \epsfig{file=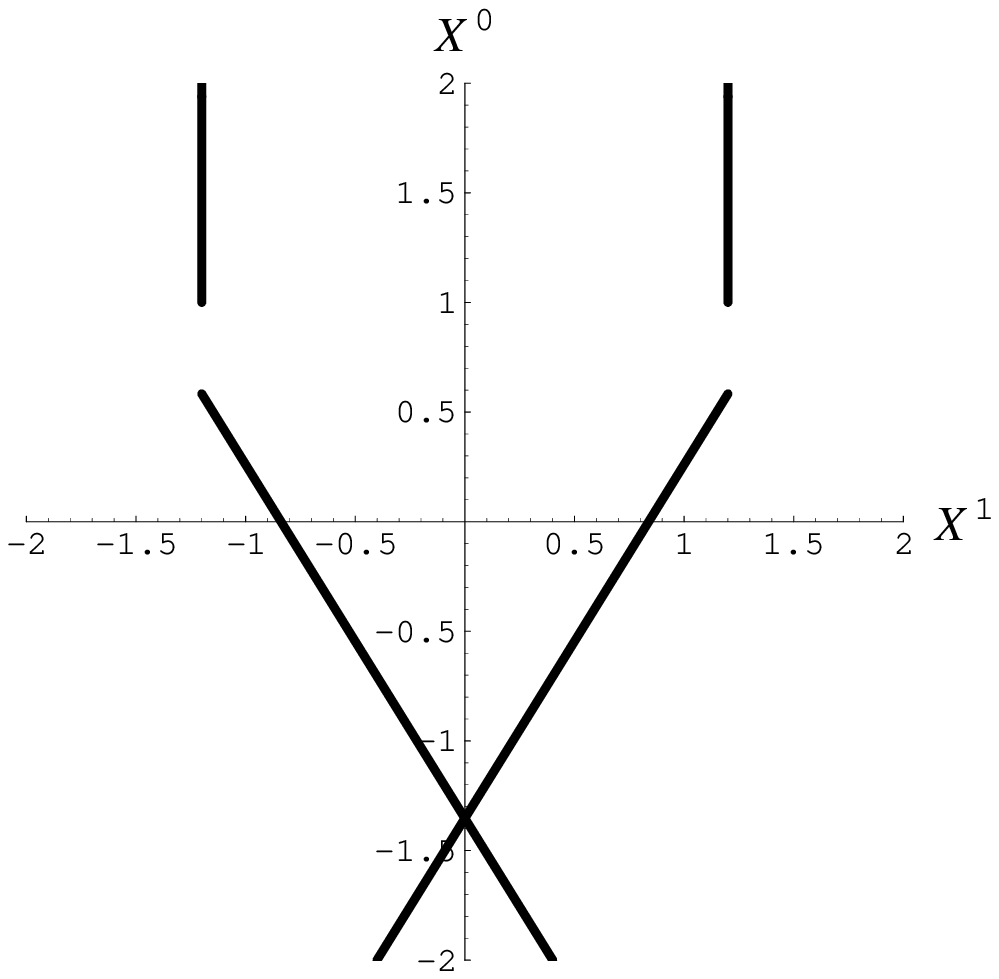, width=2.5in}
\caption{{\bf Left:} Null geodesics (thick black lines) are shown
projected into the $Z,X^0$ plane.  The geodesics experience a
discontinuity when they reach the plane $V=V_0$ of the
Aichelburg-Sexl wave, shown as a thin line.  The dashed lines are
the $U = 0$ and $V = 0$ planes. In this example, the velocity
component $dZ/dX^0$ of the geodesic is positive both before and
after the shift, but in general for $V < V_0$ the velocity $dZ/dX^0$
may be positive, negative, or zero depending on $b$.  {\bf Right:}
Null geodesics are shown projected into the $X^1,X^0$ plane.   Below
$V=V_0$, the geodesics focus toward $X^1 = 0$. Both figures show $b
= 1.2$, $c_d \mu = 1$, $V_0 = 2$, and $d= 5$.   \vspace*{7mm}}
 \label{ztplane}}
 %%\end{figure}

We consider the general case of $d \geq 4$ spacetime dimensions. In
light-cone coordinates $V = X^0+Z$, $U = X^0-Z$ the Aichelburg-Sexl
metric \cite{AS} takes the form
 \begin{equation}
 \label{asmetric}
 ds^2
= -dU dV + d\rho^2+\rho^2 d\Omega^2_{d-3}+ \mu\Phi(\rho)
\delta(V-V_0) dV^2 \,,
 \end{equation}
 where we have defined  $\rho^2 =
(X^1)^{2}+\ldots+(X^{d-2})^2$. Here the incident mass moves in the
negative $Z$ direction at $\rho=0$ within the null plane $V = V_0$.
Note that, due to the delta-function above, the metric differs from
flat Minkowski space only within this null plane.  The result
(\ref{asmetric}) can be obtained from the mass $m$ Schwarzschild
metric in $d$ dimensions by Lorentz transforming to a frame in which
the particle travels with speed $v$.  In units where $c=1$, one then
takes the limits $m \to 0$ and $v \to 1$ in such a way that the
quantity $\mu = m (1-v^2)^{-1/2}$ appearing in (\ref{asmetric})
remains constant.

The Aichelburg-Sexl potential\footnote{In contrast to the usual
convention, we have separated the factor of $\mu$ from the rest of
the potential in order to make $\Phi(\rho)$ boost-invariant.}
appearing in (\ref{asmetric}) is given by
 \begin{equation}
 \label{aspot} \Phi(\rho)
=  \left\{
 \begin{array}{ll} -c_4  \ln\left(\frac{\rho}{\rho_0}\right)&
\textrm{if $d = 4$} \\\frac{c_d }{(d-4) \rho^{d-4}} & \textrm{if $d
> 4$}
 \end{array} \ \ ,\right.
 \end{equation}
where
 \begin{equation}
 c_d = \frac{16 \pi G_d}{\Omega_{d-3}} \,  ,
 \end{equation}
 \noindent
$\Omega_n$ denotes the volume of $S^n$, and  the length scale
$\rho_0$ is an arbitrary gauge choice.   Note that the potential
$\Phi(\rho)$ solves Poisson's equation in $d-2$ dimensions with a
point source.

The Aichelburg-Sexl solution will accurately describe the metric for
$\rho \ge r$, where $r$ is the transverse size of the incident
object. The metric for $\rho < r$ depends on the internal structure
of the object; we will ignore the details in this region and treat
$\rho \sim r$ only as a cut-off on the singularity of $\Phi$. To
describe a beam of incident objects, this point source would be
replaced by a line, or by a line segment for a pulsed beam. We
consider here only the point source case (\ref{aspot}), or
equivalently the limit of short pulse duration $\Delta t  \ll 1/T$.

In the Rindler approximation, the black hole time-translation is the
boost generator $\chi = 2 \pi T (V\partial_V - U\partial_U)$, where
we have used the fact that the so-called surface gravity $\kappa$
which determines the normalization of the boost generator is related
to the black hole temperature through $\kappa = 2 \pi T$ in units
with $\hbar =c=1$. Thus the energy of the incident object  is
 \begin{equation}
 E = - \chi_\nu p^\nu = 2 \pi \mu T V_0.
 \end{equation}

We wish to use the above metric to locate the (future) black hole
event horizon, which as always  is determined by a boundary
condition in the far future.  Since (\ref{asmetric}) is precisely
the Minkowski metric for $V > V_0$, the horizon there must agree
with some null plane; i.e., with a Rindler horizon.  We take this
null plane to be $U=0$. The past horizon of the black hole is
similarly determined by a boundary condition in the far past; we
take this horizon to lie at $V=0$. Having established the boundary
conditions, we need only trace the relevant null geodesics back from
the far future through the shock wave at $V=V_0$ to find the full
horizon. We may think of this as calculating the deviation of the
true event horizon from the unperturbed event horizon at $U=0$.

The geodesic equation
 \begin{equation}
 \label{geo}
 \frac{d^2
x^\nu}{d\lambda^2}+\Gamma^{\nu}_{\rho \sigma}\frac{d
x^\rho}{d\lambda} \frac{d x^\sigma}{d\lambda} = 0
\end{equation}
was studied for (\ref{asmetric}) in \cite{FPV} and it is instructive
to review their calculation.   Setting the index $\nu = V$ in
(\ref{geo}) reveals that $V$ is an affine parameter, so henceforth
we will set $\lambda = V$. Setting $\nu = \rho$ then yields
 \begin{equation}
 \frac{d^2 \rho}{d V^2}-\frac{\mu}{2} \frac{d \Phi}{d \rho} \,\delta(V-V_0) =
 0,
 \end{equation}
while $\nu = U$ leads to
 \begin{equation}
  \frac{d^2 U}{d V^2}- \mu \Phi(\rho) \frac{d}{dV} \delta(V - V_0) -2\mu  \frac{d\Phi}{d \rho} \frac{d\rho}{d V} \, \delta(V - V_0) = 0\,.
  \end{equation}
These equations are solved by
 \begin{eqnarray}
  \label{rhosol} \rho(V) &=& b+ \mu \frac{\Phi'(b)}{2} \left(V_0 - V\right) \Theta(V_0 -
  V) , \\
  \label{usol} U(V) &=& \mu^2 \frac{\Phi'(b)^2}{4} (V-V_0) \Theta(V_0 -V) - \mu \Phi(b) \Theta(V_0 - V) \, ,
  \end{eqnarray}
where we have chosen boundary conditions such that far in the future
(i.e., $V > V_0$) we have $\rho \to b$, $U \to 0$ (see
Fig.~\ref{ztplane}).  It is straightforward to check that $ds^2 = 0$
along such curves, so these are indeed null geodesics.

 %\begin{figure}%\begin{center}%\includegraphics[width=3.5in]{caustics.eps}%\end{center}
 \FIGURE{\epsfig{ file=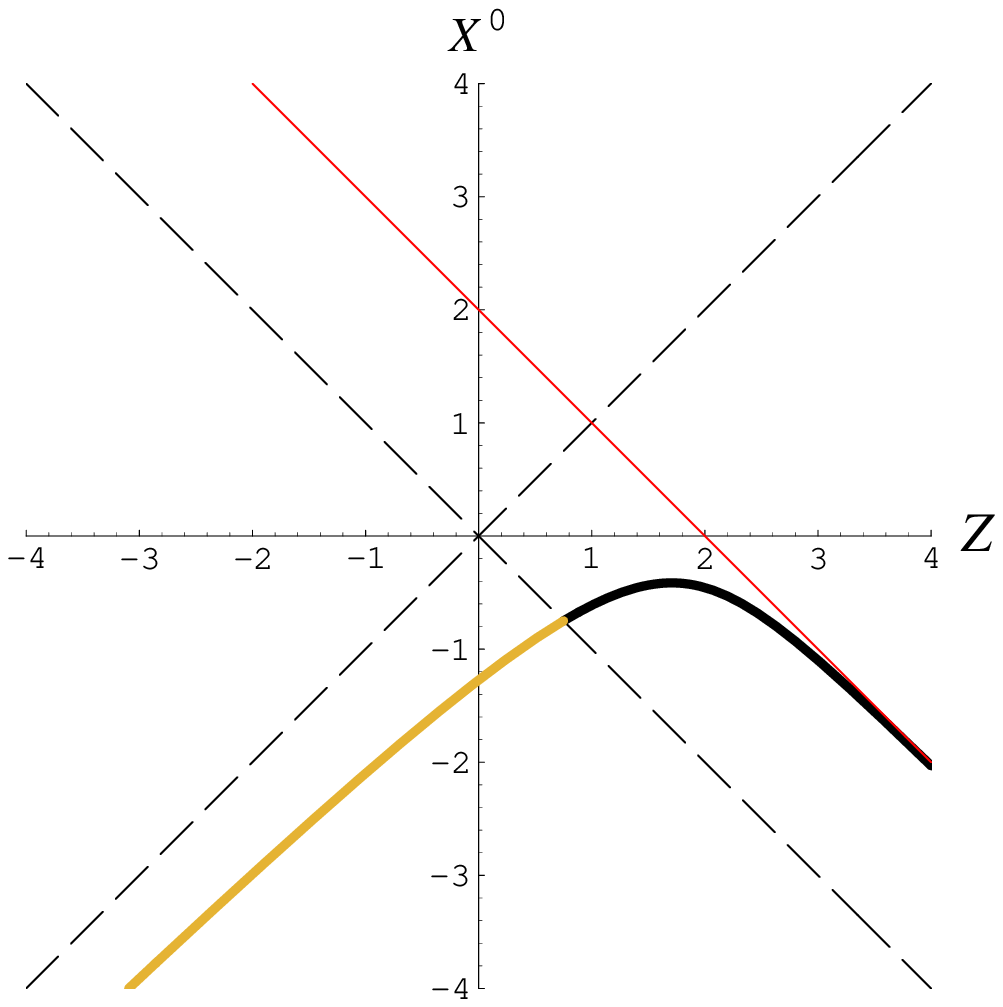,width=2.2in}
 \caption{The caustic is shown in the $Z,X^0$ plane
for $c_d \mu = 1$, $V_0 = 2$, and $d =5$.  For $b> b_*$, the caustic
is hidden behind the past horizon. For $b< b_*$ (black segment) the
caustic lies above the past horizon. This portion of the caustic
nucleates the ``arm" of the black hole.
 \vspace*{4mm}}
 \label{caustics}}
 %\end{figure}

For $V > V_0$, the geodesics are straight lines of constant $\rho$
in the $U=0$ plane.  When a geodesic of given impact parameter $\rho
= b$ reaches $V = V_0$, there is a discontinuous jump\footnote{ A
$\rho$-independent relative shift of the regions above and below
$V=V_0$ is a gauge transformation which changes $\Delta U$ by a
constant. It is in this sense that $\rho_0$ in (\ref{aspot}) is a
gauge parameter for $d=4$. In contrast, the $\rho$-dependent part of
$\Delta U$ is a physical effect \cite{DtH} describing the integrated
relative shift of geodesics passing through the shock wave. Were the
delta function in (\ref{asmetric}) replaced by a smooth function of
compact support, the geodesics would become continuous but would
still experience a net displacement $\Delta U$ as they traverse the
shock.}  in $U$ given by $\Delta U = - \mu \Phi(b),$ where the sign
corresponds to tracing the geodesics backwards from $V = \infty$.
For $d> 4$, we have $\Delta U < 0$, and $\Delta U \to 0$ as $b \to
\infty$. This is not the case for $d = 4$ due to a logarithmic
divergence.

Below the discontinuity our null geodesics are once again straight
lines, but since
 \begin{equation}
 \frac{d \rho}{d X^0} = -  \mu \frac{\Phi'(b)}{1+ \frac{\mu^2\Phi'(b)^2}{4}}> 0 \qquad \textrm{for $V <
 V_0$} ,
 \end{equation}
 they now
 begin to
 focus toward $\rho = 0$ (Fig.~\ref{ztplane}).
All geodesics with impact parameter $b$ intersect at a common point,
and the set of such intersections traces out a caustic
 as a function of $b$.
Physically, this caustic describes the nucleation of the event
horizon.

From (\ref{rhosol}) and (\ref{usol}) we see that the caustic occurs
at
 \begin{equation}
 \label{vc}
  V_c = V_0 +\frac{2 b}{\mu \Phi'(b)}
  \end{equation}
  \begin{equation}
  \label{uc}
   U_c = -\mu \Phi(b)+\frac{b\mu}{2} \Phi'(b) \,.
   \end{equation}
This curve is plotted in Fig.~\ref{caustics}.    We have
distinguished the $V < 0$ and $V > 0$ parts of the caustic, as only
the $V > 0$ part of the spacetime is physically relevant; the rest
is hidden behind the past horizon. Since geodesics with large $b$
emerge from the past horizon, we will refer to them as forming the
``body'' of the event horizon, while geodesics with small $b$
intersect the caustic  at $V >0$ and form the ``arm'' of the horizon
described in the introduction.  We see from Fig.~\ref{caustics} that
the caustic along which the arm nucleates approaches the world line
of the incident object in the limit $b\rightarrow 0$. Thus the arm
reaches out in the direction from which the object is incident by an
amount that depends on the cut-off $r$.

\subsection{Extraction of parameters}\label{extract}
%\begin{figure}%\begin{center}%\includegraphics[width=2.5in]{horint.eps}%\end{center}
\FIGURE[t]{\epsfig{file=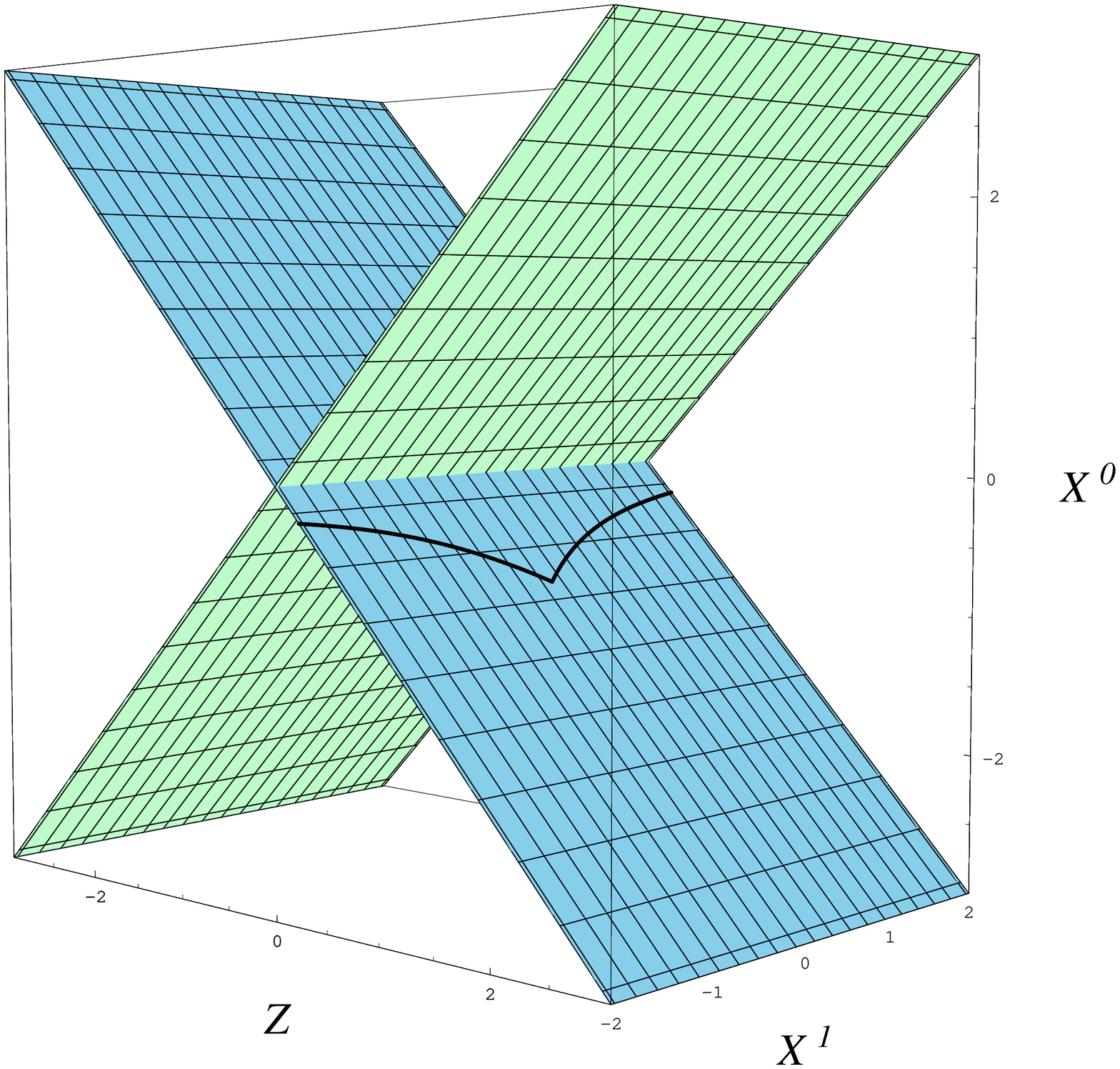, width=2.5in}
\epsfig{file=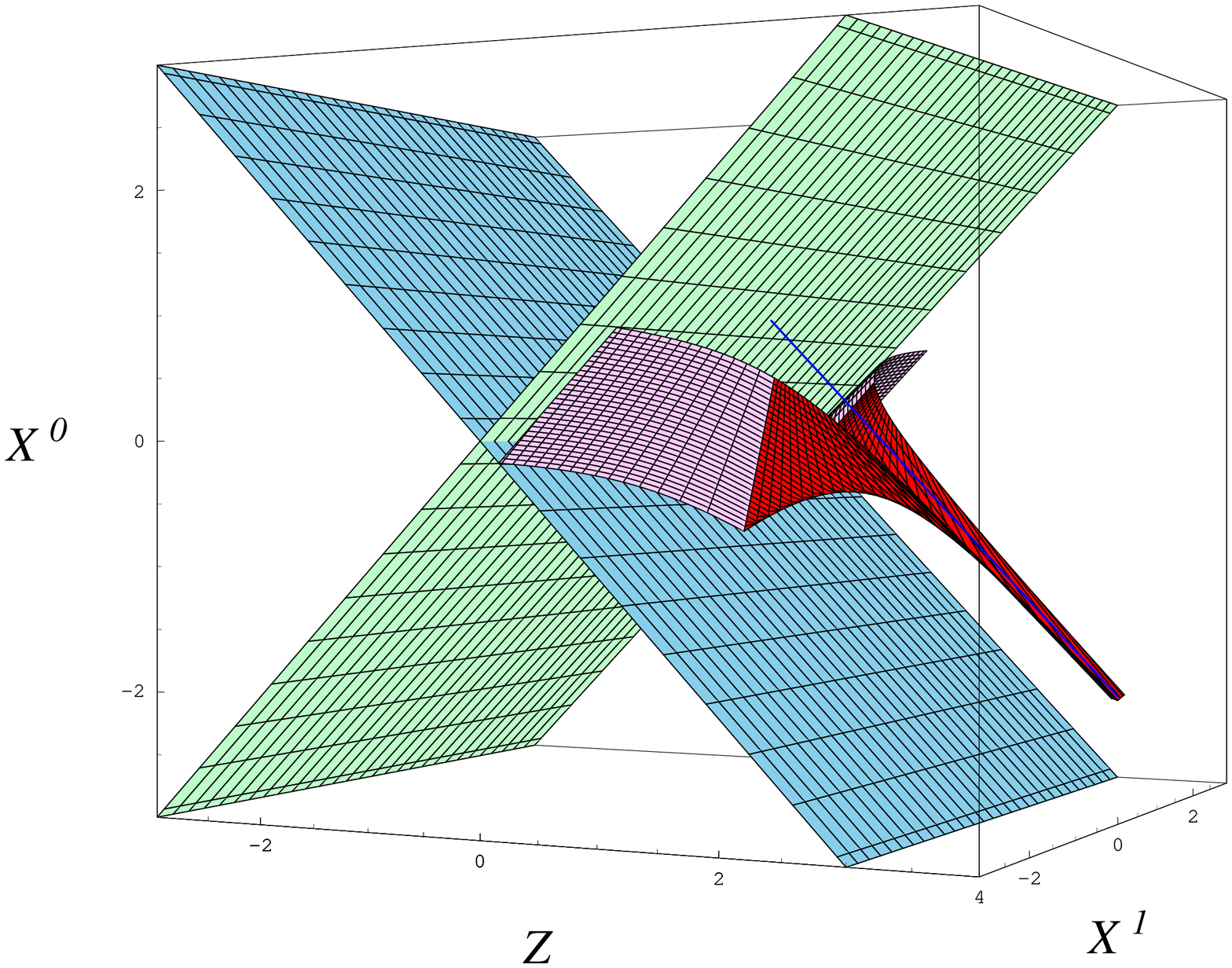, width=3in}
 \caption{{\bf Left:} The distortion of the body of the horizon by the
gravitational field of the incoming object.  The curve shows where
geodesics with $b> b_*$ intersect the past horizon.   The cusp
corresponds to $b = b_*$, where the caustic intersects the past
horizon. {\bf Right:}  The body ($b
> b_*$, lighter part of the curved surface) and arm ($b
< b_*$, darker part of the curved surface) of the horizon. The
surface shown is generated by geodesics for a range of impact
parameters $b$.  The portion of the surface above the caustic and
with $V < V_0$ is shown.  For small $b$ the spacelike curve of
caustics from which the arm nucleates is seen to approach the line
($V = V_0$, $X^i = 0$) showing the path of the incoming particle.
Both figures show $c_d \mu = 1$, $V_0 = 2$, and $d = 5$.
 }
 \label{horint}
 \label{surface}}
 %\end{figure}
Within our approximations, the above equations determine the shape
of the horizon and the newly-nucleated arm.  We now extract
parameters describing the various length scales involved for use in
our discussion of deconfined plasmas.  Since both body and arm are
dynamical objects, their lengths (and widths) will be functions of
time or, more generally, of the spacetime surface on which they are
measured. We will make certain natural choices below.

Let us begin with the body of the future event horizon, which
consists of geodesics that emerge from the past horizon.   The point
where a given geodesic intersects the unperturbed past horizon is
given by setting $V = 0$ in (\ref{rhosol}) and (\ref{usol}). The
corresponding curve is plotted in Fig.~\ref{horint} (left), and
shows the distortion (at the past horizon) of the black hole body by
the incoming flux of energy. For $d > 4$, geodesics with large $b$
experience only a small deflection, and the body at large $b$
approaches the unperturbed event horizon.  The cusp in
Fig.~\ref{horint} (left) is the event at which the caustic emerges
from the past horizon. It is described by $V_c = 0$ in (\ref{vc})
and thus by impact parameter\footnote{The result $b_*
\sim(E/T)^{\frac{1}{d-2}}$ is quite general and does not depend on
the use of the Aichelburg-Sexl approximation.  See \cite{AMV}.}
 \begin{equation}
 b_* = \left( \frac{4 G_d E }{\Omega_{d-3} T} \right)^{\frac{1}{d-2} } \, .
 \end{equation}
Both the evolution of the body and the nucleation of the arm can be
seen (Fig.~\ref{surface} (right)) by plotting the geodesics for many
values of $b$.

We now extract some relevant length scales describing the
deformation of the unperturbed horizon, including the length of the
arm. One useful measure of distance is the proper distance from the
surface $U=0, V=0$ where the unperturbed horizons intersect.
Surfaces of constant proper distance contain the worldlines of
static observers outside the unperturbed black hole. These surfaces
are just hyperbolae in the Minkowski coordinates $(X^0, Z)$
satisfying
 \begin{equation}
 Z^2 - (X^0)^2 = -U V = a^2
 \end{equation}
for some constant $a$.

We can ask which of these hyperbolae intersect a geodesic with given
impact parameter $b$, though we must take some care due to the jump
across the Aichelburg-Sexl plane. Let $P$ denote the point on the
geodesic immediately below the shift, so that  $U(P) = -\mu
\Phi(b)$, $V(P) = V_0$.  There are two cases: In the first case,
$dZ/dX^0$ at $P$ is greater for the geodesic than for a surface of
constant $a$.  In this case the geodesic is at $a < a(P)$ for $V <
V_0$ and we may say that this part of the horizon extends precisely
to $a(P)$. This occurs at large $b$, say $b > \hat b$.  In the
second case $(b < \hat b)$, one finds that $dZ/dX^0$ at $P$ is less
for the geodesic than for a surface of constant $a$. In this case,
the extension of the horizon in the $Z$ direction is characterized
by the value $a$ associated with the hyperbola tangent to the
geodesic below $P$. The two cases are shown in Fig.~\ref{hyper}.
 %\begin{figure}%\begin{center}%\includegraphics[width=3.5in]{hyper.eps}%\end{center}
 \FIGURE{\epsfig{file=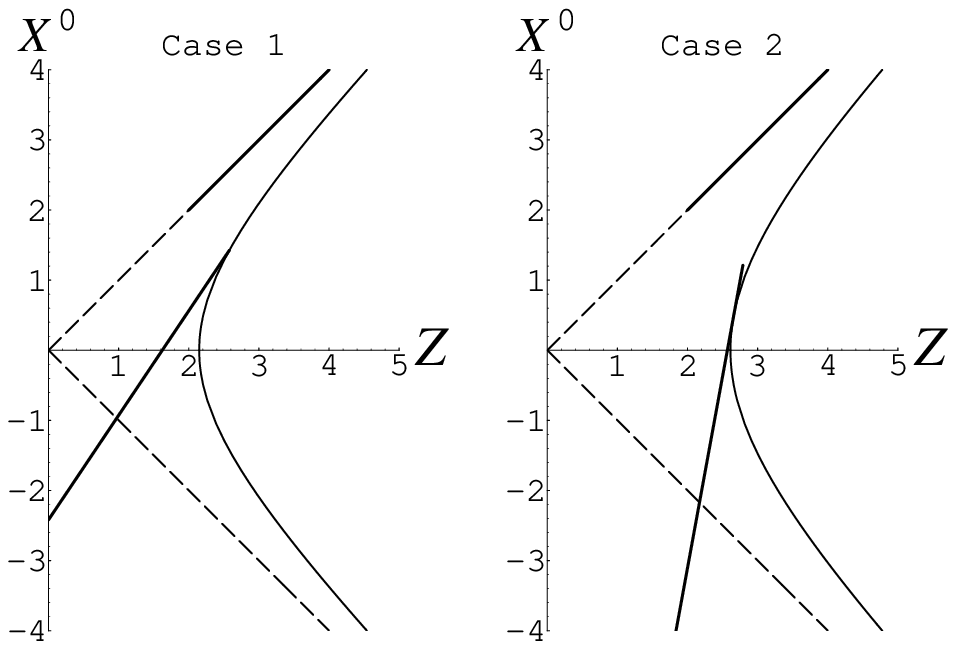,width=3.5in}
 \caption{Null geodesics and some hyperbolae for constant $a$ are shown for $c_d = 1.5$, $V_0 = 4$, and $d = 5$.
 Case 1: For
$b$ sufficiently large (e.g., $b=1.3$), the desired hyperbola
intersects the geodesic at the point $P$, that is, immediately after
the shift. Case 2: If $b$ is smaller than some critical value (e.g.,
for $b=0.95$), the geodesic is tangent to the desired hyperbola at a
point with $V < V_0$.}
 \label{hyper}}
 %\end{figure}

We restrict our analysis to the case $d > 4$, where the perturbed
horizon asymptotically approaches $U=0$ so that there is a
well-defined asymptotic bifurcation surface from which to measure
proper distance.    The results are:
 \begin{eqnarray}
 \label{hyplength}
  {\rm Case \  1:  \ \ For} \ b > \hat b \quad \, a &=& a(P)=
  \sqrt{ \frac{ 8 G_d E}  {(d-4) \Omega_{d-3} b^{d-4}   T } } ,  \cr
  {\rm Case \  2:  \ \ For}\  b <\hat b  \ \ \ \ a
  &=& \frac{ b}{d-4} + \frac{ 2  G_d E }{ \Omega_{d-3} T b^{d-3} }   .
  \
  \end{eqnarray}
where $\hat b$ is given by
 \begin{equation}
 \label{bhyp}
 \hat b = \left(\frac{d-4}{2} \right)^{\frac{1}{d-2}} b_*
 =  \left(\frac{d-4}{2} \right)^{\frac{1}{d-2}} \left( \frac{4 G_d E }{\Omega_{d-3} T}
 \right)^{\frac{1}{d-2} } .
 \end{equation}

The maximum extension  (or length) of the arm\footnote{In the regime
of most interest, $r \ll b_*$, the contribution from the black hole
body is negligible in comparison to the length of the black hole arm, so
henceforth we will refer to the ``maximum extension of the arm'' as
simply the ``length of the arm.''} is just (\ref{hyplength})
evaluated at the cut-off $b = r$. Note that for $d \ge 6$ any
geodesic intersecting the arm satisfies $b \le b_* \le \hat b$, so
that case 2 always applies. Even for $d=5$, case 2 applies whenever
the effect is significant, $r \ll b_*$. In this limit one finds
 \begin{equation}
 \label{BHlength}
 L_{BH \ arm} \approx \frac{2 G_d E}{\Omega_{d-3} T
 r^{d-3}}.
 \end{equation}
We note once again that, due to the Rindler approximation, this
result is justified only in cases where  $L_{BH \ arm}$ is smaller
than the characteristic curvature scale of the spacetime near the
horizon.

We may also characterize the arm by its width in the Aichelburg-Sexl
plane. At each $U$ this width is just the impact parameter, which
ranges from $b_* = \left( \frac{4 G_d E }{\Omega_{d-3} T}
\right)^{\frac{1}{d-2} }$, where the arm connects to the body, to
the cut-off $r$ at the tip.

\section{Arm nucleation for plasma balls}
\label{pbs}

In gauge/gravity duality, balls of deconfined plasma surrounded by a
confining phase are dual to black holes that are localized in the
gauge theory directions.  Such black holes naturally have
temperature $T=T_d$, where $T_d$ is the temperature of the
deconfinement phase transition \cite{AMW}. A particle incident on
the gauge theory plasma ball corresponds to a flux of energy
incident on the black hole.  If this flux is large enough, the black
hole may nucleate an arm as described in section \ref{geodesics}.

 % \begin{figure} \begin{center}  \includegraphics[width=2.5in]{plasma.eps} \end{center}
 \FIGURE{\epsfig{file=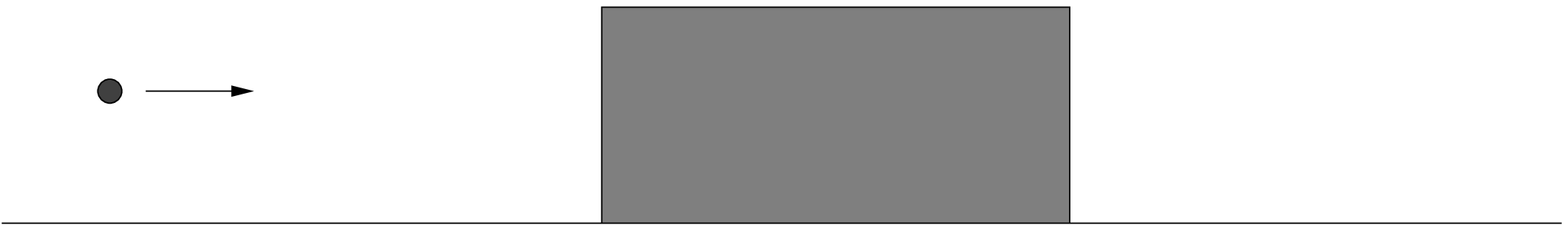,width=2.5in}
 \caption{A particle incident on a black hole dual to a ball of deconfined plasma.}
 \label{plasmafig} }
 % \end{figure}

How precisely should we understand the corresponding effect in the
gauge theory? In some sense the plasma ball must also nucleate an
arm along the path of the incoming particle.  What is interesting is
that Fig.~\ref{surface} (right) indicates that this nucleation
occurs {\it in front} of the incoming particle; i.e., before the
particle arrives.

Such behavior may appear to be acausal, but the same words might be
applied to the nucleation of the arm on the black hole.  Yet the
physics of the black hole arm is clear.  Gravity propagates
causally, and the spacetime in the region of the arm has no locally
measurable differences from the spacetime outside.  It is the fact
that one traces the event horizon backwards from the far future that
allows the arm to form before the particles arrive.  In other words,
our designation of the arm as being ``part'' of the black hole
describes the fact that, due to the imminent arrival of the incident
energy, the gravitational interactions {\it will} become large
enough that no information from the arm-like region will be able to
escape to infinity.

This suggests a straightforward interpretation of the corresponding
``arm'' of the plasma ball:  Due to causality, the arm is locally
indistinguishable from the confining vacuum.  In particular, its
energy density remains that of the vacuum and the arm contains no
plasma.  Instead,  the arm marks a region of spacetime in which any
additional particle has a large probability to be caught up in the
interactions that {\it will} ensue and thus to fuse with the plasma
ball.  In effect, the arm describes a temporary and highly
anisotropic enlargement of the capture cross-section associated with
the plasma ball.  To remind the reader of this interpretation, we
will henceforth refer to such arms as ``virtual'' in the context of
plasma balls.

 \FIGURE[h!]{\epsfig{file=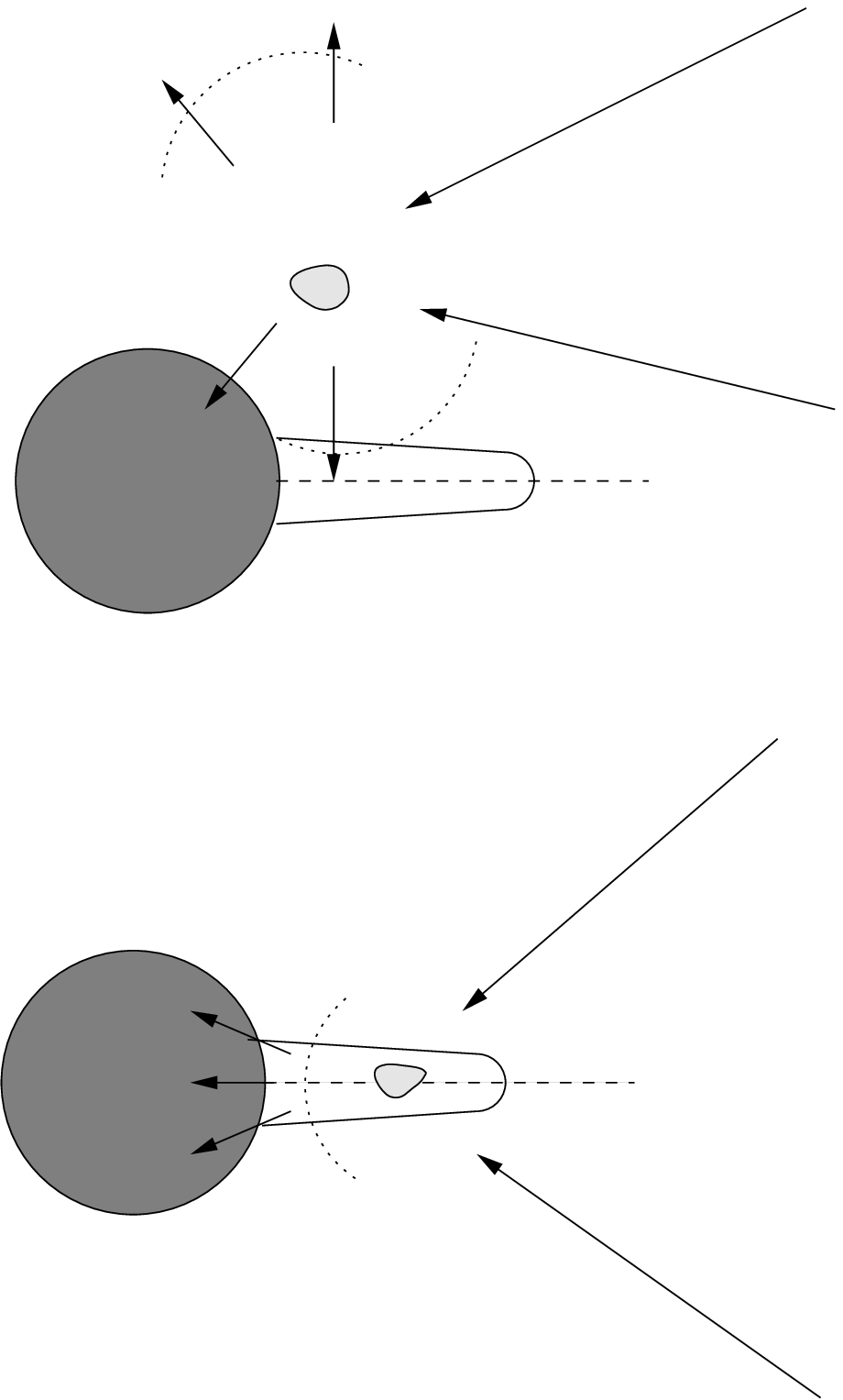,width=2.5in}
\caption{{\bf Top:} Particles produced in a collision exterior to
the virtual arm can reach the detector.  {\bf Bottom:}  Particles
produced in a collision inside the virtual arm fuse with the plasma
ball and will not be detected far away.}
 \label{newplasmafig} }
Now, even without the nucleation of virtual arms,  the capture
cross-section for particles incident from infinity will be somewhat
larger than the size of the plasma ball itself. Recall, for example,
that while the event horizon for the familiar static Schwarzschild
black hole lies at area-radius $2G_4 M$, any free particle passing
within the sphere of area-radius $3G_4 M$ is necessarily captured by
the black hole. It is thus clear that there will be a similar effect
in the gauge theory dual\footnote{Our Rindler-space near-horizon
approximation is enough to predict that any low-mass particle
incident from infinity is captured as long as it approaches within
some distance of order the curvature scale of the spacetime near the
horizon. However, the full capture cross-section may well be even
larger.}.

However, the arm describes a region of spacetime tied even more
strongly to the plasma ball. Suppose that two additional (probe)
beams are aimed so as to collide near the plasma.  If the
interaction region lies outside the virtual arm, then some of the
debris from the interaction can escape to infinity.  However, when
an interaction takes place inside the virtual arm, {\it all} of the
particles produced in the collision will have large amplitudes to
fuse with the plasma ball and no signal will reach an external
detector.   This difference is illustrated in
Fig.~\ref{newplasmafig}.

One may hope to one day observe the effects of similar virtual arms
in experiments with deconfined QCD plasmas.  In this context, recall
that our description of black hole arms in section \ref{geodesics}
is universal in the limit of {\it i}) high velocity collisions and
{\it ii}) studying the region near the black hole.  As a result,
experiments that probe plasmas in the corresponding regime can
provide a new universal test of the extent to which such plasmas
admit a dual description by {\it any} semi-classical gravity system,
without first needing to propose a particular dual theory.

To illustrate this idea, let us consider the length  of the arm
(\ref{BHlength}), which describes the limit of  large velocity for
the incident particles.  We wish to translate this result into a
length $L_{plasma \ arm}$ for the arm of virtual plasma measured in
gauge theory units.  Since (\ref{BHlength}) contains a factor of
$G_d$, and since the relation between $G_d$ and the gauge theory
parameters is model-dependent, the result will no longer be
completely universal. However, the scaling of $L_{plasma \ arm}$
with various parameters describing the incident particle will be
universal.  Precise predictions can of course be obtained for
particular models; we give an example at the end of this section.

The main task is to determine how distances (such as $r$, $L_{BH \
arm}$) in the gravity description are related to gauge theory
distances. Here the high symmetry of our near-horizon limit is quite
useful.  Recall that we consider regions close enough to the horizon
to regard the surface as flat.  Thus we have Euclidean symmetry
along the horizon as well as translation symmetry in time. There is
also a reflection symmetry which inverts any constant $z$ surface
through any point on that surface.   As a result, in terms of the
coordinates $\rho, z, t$ the gauge theory metric must be of the form

 \begin{equation} ds^2_{YM}  = \sigma^2(z) \left( -  f(z) dt^2 + g(z) dz^2 + d
 \rho^2 + \rho^2 d\Omega_{D-3}^2 \right)\,,
 \end{equation}
where we have taken the gauge theory to live in $D$ dimensions.  In
addition, the original coordinate $t$ was normalized to the unit
time-translation in the gauge theory.  Thus, $f = \sigma^{-2}$.
Since the gauge theory space-translation (say, in the transverse
directions) also has constant norm, $\sigma$ must be a constant. The
gauge theory metric is thus
 \begin{equation} ds^2_{YM}  =  -   dt^2 +  \sigma^2 \left( g(z) dz^2 + d
 \rho^2 + \rho^2 d\Omega_{ D-3}^2 \right).
 \label{gtm}
 \end{equation}
Any such metric describes flat Minkowski space (as desired), but the
particular form of (\ref{gtm}) tells us how to translate our bulk
coordinates into gauge theory terms.

Information about $g(z)$ may now be obtained by considering the
energy of strings stretched in both gauge theory and bulk. First
recall that (\ref{BHlength}) was obtained in the limit where one is
close enough to the black hole horizon to approximate the metric by
the Rindler metric,
 \be
  ds^2 = - (2\pi T_d)^2 z^2 dt^2 + dz^2
+ d\rho^2 +\rho ^2 d\Omega^2_{d-3}\, .
 \ee
Now consider a bit of string located near the horizon and oriented
parallel to the horizon,  that is, at constant $z$.  If, for
example, such a string were stretched in the $\rho$ direction (at
fixed angles) with coordinate extent $\Delta \rho$, then its energy
would be $E = \ell_s^{-2} T_d z \Delta \rho$. Far from the plasma
ball, the energy of the corresponding stretched string in the gauge
theory would be $E = (2 \pi \alpha')^{-1}  (\sigma \Delta \rho)$
where $\alpha'$ is the gauge theory Regge slope.  However, these two
expressions cannot be equal as the gauge theory expression does not
depend on $z$.  The point here is that there is an attractive force
between the string and the plasma ball, so that proximity to the
plasma ball changes the effective string tension (and thus the
effective Regge slope) by a $z$-dependent shift in the tension.

Let us call the effective Regge slope $\beta(z) \alpha'$. Setting
the resulting energies equal yields $\frac{\sigma}{ \beta}= \frac{2
\pi \alpha' T_d z} {\ell_s^2}$.  If we assume that the
renormalization factor $\beta$ depends only on $z$ and not on the
orientation of the string, we may obtain information about $g(z)$ by
considering bits of string stretched in the $z$-direction:
 \begin{equation}
 \ell_s^{-2} T_d z \Delta  z = E = \frac{1}{2 \pi  \beta  \alpha'} \, \sigma (\sqrt{g} \Delta z),  \ \ {\rm or}  \ \ g = 1.
 \end{equation}

This implies that spatial distances are merely rescaled by a factor
of $\sigma$ when passing from gravity to gauge theory measurements.
Thus, the size of the incident object in the gauge theory  is
$\tilde r = \sigma r$ and the length of the virtual plasma arm is
\begin{equation}
\label{PBlength}
 L_{plasma \ arm} \approx \frac{2 \hat G_d \hat \sigma^{d-2} \alpha'{}^{\frac{d-2}{2}}}{N_c^2 \Omega_{d-3} T_d}
 %\left(
\frac{E}{ {\tilde r}^{d-3}}
%\right)
,
\end{equation}
where $\hat G_d$ and $\hat \sigma$ are defined by $\hat G_d  = N_c^2
G_d/\ell_s^{d-2}$  and $\hat \sigma =  \sigma \left(
\frac{\ell_s^2}{\alpha'} \right)^{1/2} .$ We conclude that the
length of the virtual arm scales with the energy of the incident
particle and inversely with a power of the particle's characteristic
size. This power determines the effective dimension of the gravity
dual.  Similarly, the width of the arm ranges from $W^{max}_{plasma
\ arm} = \hat \sigma \alpha'{}^{1/2} \left( \frac{4 \hat G_d
E}{N_c^2 \Omega_{d-3} T_d } \right)^\frac{1}{d-2}$ where it connects
with  the plasma ball to $W^{min}_{plasma \ arm} = \tilde r$ at the
tip.

It is clear that $\hat G_d$ is a dimensionless model-dependent
function; i.e., a function of $\lambda$.  To clarify the nature of
$\hat \sigma$, let us first consider $\sigma =  2 \pi \beta  \alpha'
T_d z/\ell_s^2$.  For a given plasma ball, we already know that
$\sigma$ is a dimensionless constant summarizing the scale of gauge
theory proper distances relative to gravity proper distances near
the black hole.  On the other hand, far from the black hole the
scale factor relating gauge theory relative to gravity distances is
just $\left( \frac{\alpha'}{\ell_s^2} \right)^{1/2}$.  Thus, $\hat
\sigma =  \sigma \left( \frac{\ell_s^2}{\alpha'} \right)^{1/2}$
gives precisely the scale factor of gravity distances near the black
hole relative to those far away; i.e., it is determined by the black
hole metric. As a property of a classical solution of a pure gravity
theory, $\hat \sigma$ cannot depend on $G_d$, and must be a function
only of $L/R$, where $R$ is the size of the black hole and $L$ is
the AdS scale.  Let us, however, consider the relation $\hat \sigma
= \left( \frac{\ell_s^2}{\alpha'} \right)^{1/2} \sigma =  2 \pi
\beta \alpha'{}^{1/2} T_d z/\ell_s $ from the gauge theory point of
view.  (In terms of the gauge theory proper distance $\tilde z =
\sigma z$ this is $\hat \sigma^2 =
 2 \pi \beta T_d \tilde z$.)  The only factor through which the size of the black hole (i.e., the size of the plasma ball)
 might enter is $\beta$, the renormalization of the Regge slope.  But since this renormalization is due to interactions with
  the plasma ball, at small $\tilde  z$ (small separation from the plasma ball) one expects $\beta$ to depend only on $\tilde z$
   and to become independent of $R$.  Thus we deduce that $\hat \sigma$ is in fact a constant of order one which depends only on qualitative
   features of the model.  From the gravity perspective this means that $\hat \sigma$ is independent of $G_d,L,R$,
   while from the gauge theory perspective $\hat \sigma$ is independent of $\lambda, N_c, \alpha'$.

A particular gauge/gravity duality of this kind was considered in
\cite{wit,AMW}.  There a (2+1) dimensional theory was obtained by
taking the low energy limit of ${\cal N} = 4$ super Yang-Mills
compactified on a Scherk-Schwarz spatial circle.  This theory is
dual to a IIB supergravity system whose ground state is the
Cartesian product of the AdS soliton \cite{HMsol} and $S^5$.   In
the $\lambda \to 0$ limit, the field theory is pure (2+1) Yang Mills
\cite{wit}.   At large $\lambda$, where the classical gravity
description is valid, the gauge theory is unfamiliar. Nevertheless,
it was conjectured in \cite{AMW} that finite mass black holes have a
dual description in this theory in terms of localized lumps of
deconfined plasma. If the incident particle has no $R$-symmetry
charge, the corresponding object in the gravity dual is delocalized
on the $S^5$ so that the effective dimension is $d=5$ and (using the
conventions of section \ref{review} below) $\hat G_5 = 2^{-1/4} \pi
\lambda^{3/4}$.

One would also like to determine the constant $\hat \sigma$ for this
model.  In principle, this can be read off from the bulk metric
which describes the domain-wall limit considered in \cite{AMW}
(where the black hole fills a half-plane).  In that case $\hat
\sigma = |k|_{\infty}/|k|_{BH}$, where $|k|_{\infty}$ is the norm of
a translational Killing field $k$ along the domain wall evaluated
far from the black hole but on the IR floor of the spacetime and
$|k|_{BH}$ is the norm of $k$ evaluated near the horizon of the
black hole. Using Fig. 6 of \cite{AMW}, it is reasonable to estimate
that $\hat \sigma \approx 1$.

\section{Localized heating of deconfined plasmas}
\label{mesons}

Section \ref{pbs} above discussed the gauge theory dual of a process
in which a high-velocity object is incident on the black hole from a
gauge theory direction. However, another interesting setting occurs
when the object is incident on the black hole from the holographic
direction. In this case, the incident object represents a localized
heating of the plasma, perhaps due to the decay of massive
quasi-particles (see section \ref{melt} below).

If the theory has additional phase transitions above the
deconfinement temperature $T_d$, then this local heating can induce
the formation of a bubble of the high temperature phase.  We now
study such bubble formation and the ensuing bubble dynamics in a
particular gauge/gravity correspondence. %involving confining gauge
%theories

\subsection{Brief review: AdS/CFT with flavor}
\label{review}

Recall \cite{KR, KK} that in AdS/CFT, a small number of flavors
($N_{f} \ll N_c$) of fundamental matter (``quarks'') may be
described by probe D-branes in a background AdS space.  The
transverse displacement of the probe branes from the D3-branes  that
describe AdS space sets the mass of the quarks. We consider the
case where the bulk solution contains a black hole at temperature
$T$. Although the probe branes are
attracted to the black hole, at high enough quark mass the tension
of the branes is sufficient to balance this attractive force and the
probe branes lie outside the horizon in what is often called a
`Minkowski' embedding \cite{MMT1,MMT2}. In this phase, the gauge
theory contains stable mesons with a discrete mass spectrum.
However, above a critical temperature $T_f$,  the gravitational
force overcomes the tension and the probe branes fall through the
horizon. The associated embeddings are referred to as `black hole'
embeddings. In this phase, the gauge theory has no stable mesons;
the mesons melt. A detailed thermodynamic analysis of these
embeddings reveals that the system undergoes a first order phase
transition at $T = T_{f}$, where the probe branes jump
discontinuously from a Minkowski to a black hole embedding
\cite{MMT1, MMT2}.

This phase transition has been studied extensively in systems of
$N_f$ probe D7-branes in the background of $N_c$ black D3-branes,
with $N_f \ll N_c$. Without the D7-branes, the system is conformal
and does not confine. Nevertheless, it can be a good model of a
confining system at sufficiently high temperatures $T \gg T_d$. We
now specialize to this model and fix the notation to be used below.

The background metric for the black D3-branes may be written
 \begin{equation}
 \label{d3metric} ds^2 = \left(\frac{u}{L}\right)^2 \left(-f dt^2
+d\vec{x \,}^2 \right)+  \left(\frac{L}{u}\right)^2
\left(\frac{du^2}{f} + u^2 d\Omega^2_5\right),
 \end{equation}
where $f = 1-(u_0/u)^4$.  The gauge theory directions are $\vec{x} =
(x^1, x^2, x^3)$, and the line element of the $S^5$ is
 \begin{equation}
 d\Omega^2_5 = d\theta^2 + \sin^2 \theta d\Omega^2_3 + \cos^2 \theta
 d\phi^2 \,.
 \end{equation}
The length scale $L$ can be expressed as
 \begin{equation}
 \label{L4}
 L^4 = 4 \pi g_s N_c \ell_s^4 = 2 \lambda \ell_s^4 \,,
 \end{equation}
where $g_s$ is the string coupling, $\ell_s$ is the string length,
and we have defined the 't Hooft coupling $\lambda = 2 \pi g_s N_c$.
The horizon is at $u = u_0$, which is related to the temperature $T$
through
 \begin{equation}
 T = \frac{u_0}{\pi L^2} \,.
 \end{equation}
As shown in \cite{MMT1,MMT2}, the critical temperature for the meson
melting phase transition is
 \begin{equation}
 T_f \sim \bar M \sim \frac{2 M_q}{\sqrt{\lambda}} \, ,
 \end{equation}
where $\bar M$ characterizes the meson mass gap at temperatures well
below the phase transition, the quark mass is $M_q \sim u_0/
\ell_s^2$, and factors of order one have been neglected.

Now, in the near horizon limit, the metric (\ref{d3metric}) can be
written in the standard Rindler form by defining the coordinates
 \begin{equation}
 \pi T z^2 = u-u_0\,, \qquad y = L \theta,
 \end{equation}
and rescaling $\vec{x} \to \frac{u_0}{L} \vec{x}$.  Then in the
limit of small $z$ and small $y$, the background metric becomes
 \begin{equation}
 \label{rindler}
  ds^2 = -(2 \pi T)^2 z^2 dt^2 + dz^2 + d\vec{x \,}^2 +
dy^2 +y^2 d\Omega_3^2 + dx_9^2 \, ,
 \end{equation}
where $\vec x = (x^1,x^2,x^3)$. For further details we refer the
reader to \cite{MMT1, MMT2, fro}.

\subsection{Bubbles of melted mesons}
\label{melt}

Consider now a thermal state of the gauge theory at some temperature
$T$ near $T_f$ but satisfying $T_f > T$. The probe branes in the
gravity dual lie on a Minkowski embedding which we may characterize
by the minimum height $z_0$ of the branes above the black hole. This
minimum height occurs at $y=0$ and is related to the constituent and
bare quark masses $M_c, M_q$ through \cite{ MMT2}
\begin{equation}
M_c \sim \frac{u_0 z_0^2}{\ell_s^2 L^2} \sim \frac{z_0^2}{L^2} M_q
\, .
\end{equation}
The constituent mass $M_c$ is the physical quark mass taking  into
account thermal corrections.

From \cite{MMT2} we see that Minkowski embeddings with $z_0 < 0.15
\, L$ are thermodynamically unstable and will transition to black
hole embeddings. Thus $z_0 \gtrsim 0.15 \, L$.   However, we wish to
use the Rindler-space approximation (\ref{rindler}), which neglects
terms in the metric of order
$$\frac{u-u_0}{u_0}  \sim \left(\frac{z}{L}\right)^2  \,.$$  Thus
for $z \sim 0.15 L$, the error is of order $(0.15)^2$. This
occurs when $T$ is close to $T_f$.  While not parametrically small,
the above error is certainly not large and we expect that the
Rindler approximation should correctly capture much of the physics.
Below, we restrict attention to such cases (which implies $M_c \sim
(0.15)^2 M_q$) and use the Rindler approximation without further comment.

Suppose that we now perturb this equilibrium by adding a massive
quasi-particle which would have been stable (or at least very long
lived) at $T=0$.  In the presence of the thermal bath all such
quasi-particles become unstable and decay, but at least at large
$\lambda,N_c$ some massive quasi-particles have lifetimes $\Delta
\tau$ much greater than their inverse masses $m^{-1}$. In the
gravity dual such resonances merely correspond to raising some
object (graviton, massive string state, or black hole) to a height
$z \gg z_0$ above the black hole horizon. We take the wave-function
of this excitation in the gauge theory directions to be a Gaussian
wave-packet centered on zero momentum with small width in momentum
space, so that any spreading of the wave-packet is very slow.  Due
to the object's large mass, such wave-packets may nevertheless also
be well-localized in position space.

Since our object was raised to a large height $z$, it quickly begins
to fall.  By the time it reaches $z \sim z_0$, the object is moving
very rapidly.  We restrict attention to an object which remains
well-localized in the gauge theory directions (i.e., a beam-like
graviton wave-packet directed downward, a massive string, or a small
black hole), so that we may approximate its effect on the probe
brane by that of an Aichelburg-Sexl shock wave of the form discussed
in section \ref{geodesics} above.

Now, there are two settings that one might consider, depending on
whether the object is localized or delocalized on the $S^5$. In the
particular model considered here, the $S^5$ has size of order $L$.
Due to various dynamical instabilities such as the Gregory-Laflamme
instability for black holes \cite{GL}, one expects stable objects
that are well-localized in the gauge theory directions and which
satisfy $r \ll L$ to also be well-localized on the $S^5$.

However, this fact means that the wavefunction of such objects will
have components with significant charge under the SO(6) $R$-symmetry
associated with this sphere.  If one eventually wishes to model more
QCD-like systems without an $R$-symmetry, then one may expect that
the case without $R$-charge provides more universal results.  Thus,
despite the fact that such solutions are unstable in our model, we
concentrate below on Aichelburg-Sexl-like solutions which are
homogeneous over the $S^5$.

In our Rindler-space near-horizon limit, this $S^5$ is merely the
${\mathbb R}^5$ spanned by $y, \Omega_{3}, x^9$.  One may think of
the resulting shock-wave solution as being described by a $d=5$
Aichelburg-Sexl solution (\ref{asmetric}) at each constant value of
$y, \Omega_{3}, x^9$.    In the notation of section \ref{geodesics}
above, one may simply take the transverse radial coordinate to be
$\rho^2 = \vec{x \,}^2$ and $d = 5$.

We found previously that null geodesics undergo a discontinuous
shift when they reach the plane of the shock. In fact, the same is
true of any worldline subject to finite forces, as such forces add
only a finite right-hand side to equation (\ref{geo}). Since the
shift diverges at $\rho =0$, it takes at least a portion of the
D7-branes' worldvolume behind the black hole horizon.  A localized
bubble of the melted meson phase has been formed.

We wish to describe the size and dynamics of this bubble.  A full
solution would require solving the D7-brane equations of motion in a
dynamic background, taking proper account of the boundary conditions
at infinity.  We can, however, make some progress within our Rindler
approximation.   Note first that, due to the scale invariance of
Minkowski space, in our Rindler approximation the curvature scale of
the initial static D7-brane at the tip $(y=0,z=z_0)$ must be set by
$z_0$ \cite{fro}.   Thus, for $y \ll z_0$ we can model the initial
configuration of the brane by the hyperplane $z=z_0$, with $x_9=0$.
In terms of the light-cone Minkowski coordinates, the brane lies on
the hyperbolic curve $-U V = z_0^2$ with $x_9=0$.

At $V = V_0$, a point on the brane at impact parameter $\rho$ will
be shifted by an amount $|\Delta U| = \mu \Phi(\rho)$. Thus,
immediately after the jump the profile of the branes is given
by\footnote{When the falling object is localized on the $S^5$ one
obtains instead
\begin{eqnarray}
\label{itube} z(\rho) &=& \sqrt{z_0^2 - \frac{E}{2 \pi T}
\Phi(\rho)} = z_0\sqrt{1-\left(\frac{R_{loc}}{\rho} \right)^6} \, ,
\cr \left(R_{loc}\right)^6 &=& \frac{4}{3 \Omega_7} \, \frac{G_{10}
E}{T z_0^2} \nonumber \, .
\end{eqnarray}
 }
\begin{equation}
\label{itube} z(\rho) = \sqrt{z_0^2 - \frac{E}{2 \pi T} \Phi(\rho)}
= z_0\sqrt{1-\frac{R_0}{\rho} } \,,
\end{equation}
where
 \begin{equation}
 \label{R0}
 R_0 = \frac{2}{ \pi} \, \frac{G_5 E}{T z_0^2} \,
 \end{equation}
and $G_5 = G_{10}/ (\pi^3 L^5)$.  We will comment further on the
fact that $R_0 \sim E$ in section \ref{disc}.

Equation (\ref{itube}) describes a cylinder of radius $R_0$ and
height $z_0$ which attaches smoothly to an infinite plane at
$z=z_0$. Our approximation is valid for $R_0 \ll z_0$, in which case
the cylinder is tall and narrow.   It therefore has a tendency to
contract.   Furthermore, as shown in the appendix, over most of our
Rindler region (for $\rho < \rho_* = (R_0z_0^2)^{1/3}/2$, or $z <
z_*$ with $z_* \approx z_0 - (R_0^2 z_0)^{1/3} $) the bubble
nucleates with $\partial \rho/\partial V < 0$.   Thus, unless other
influences from beyond $z_*$ can arrive in time to prevent it, this
part of the tube will collapse.

We now show that no such influences can arrive in time to
prevent collapse.  Some care is required due to the fact that our
initial data is given on the surface $V=V_0, z = z(\rho)$, which is
not a constant time surface in any natural coordinate system.  To
begin, consider a slice of the brane at constant $z$. If no
influences arrive to slow the collapse, this part of the brane will
collapse at $V = V_0 + \Delta V$ for
\begin{equation}
\Delta V = \rho \left(-\frac{\partial \rho}{\partial
V}\right)_z^{-1} =  \frac{V_0}{(\rho_*/\rho)^3-1} \,, \qquad \rho <
\rho_* \,,
\end{equation}
where in the last step we have used results from the appendix.
Recalling that $U = -z^2/V$ we see that an event at which a point on
the cylinder collapses to zero size has spacetime coordinates $(V,
U, \rho) \sim (V_0+\Delta V, - z_0^2/V_0+\Delta V z_0^2/V_0^2 , 0)$,
where we have assumed $\Delta V/V_0 \ll 1$.  On the other hand, the
event at $V = V_0$ where $\partial \rho /\partial V$ becomes
positive has spacetime coordinates $(V_* = V_0, U_* = -z_*^2/V_*,
\rho_*) \sim (V_0, -z_0^2/V_0, \rho_*)$.  The spacetime interval
between these events is therefore
\begin{equation}
\Delta s^2 = -\Delta U \Delta V + (\Delta \rho)^2 \sim -\left(
\frac{z_0}{(\rho_*/\rho)^3-1} \right)^2+\rho_*^2 \,,
\end{equation}
which is positive for $\rho < \bar \rho \sim \frac{1}{2^{4/3}}
R_0^{4/9} z_0^{5/9}$.  Thus, the tube collapses at least out to
height $\bar z = z_0 \sqrt{1 - \frac{R_0}{\bar \rho} } \approx z_0
\left( 1 - 2^{1/3} \left( \frac{R_0}{z_0} \right)^{5/9} \right)$.

It is interesting to ask about the rate at which the bubble
collapses.  However, this question is complicated by several
considerations associated with the fact that the bubble is far from
even local thermodynamic equilibrium.  First, the bubble does
not form at a single value of the gauge theory time $t$. Instead, it
forms on the null surface $V = V_0$. Second, there is some ambiguity
as to what one wishes to call the ``size" of the bubble. The naive
bubble size is $R_0$, the radius of the cylinder of D7-brane
intersecting the horizon at $V=V_0$. However, we have seen that the
rather larger region of the brane out to $\rho = \bar \rho \gg R_0$
eventually collapses.  A causality calculation similar to the one
above then shows that any excitation of the brane (e.g., a meson)
which begins at $\rho < \hat \rho \sim 2 (R_0^2 z_0)^{1/3}$ is
necessarily caught in the collapsing bubble, so that it is the scale
$\hat \rho$ which governs the region in which meson propagation is
highly disrupted. In terms of the gravity side of the gauge/gravity
correspondence, we will tentatively describe these effects as a
bubble of proper size $R_0$ surrounded by a rather larger
region in which propagation of disturbances on the D7-brane is highly disrupted
 (though we again emphasize that in reality the entire region is far from any local thermodynamic equilibrium).
  Using the metric  (\ref{d3metric}) to translate distances to the gauge theory, for small $R_0/z_0$ we have a
   bubble of size  $\tilde R_0 = R_0/(\pi T L) \sim  E/( \pi (0.15)^2 N_c^2T^2)$,
   surrounded by a large halo that interferes with meson propagation.

On the other hand, the time interval $\Delta t$ between the creation
and collapse of the bubble is a well-defined quantity. Using $V = z
e^{2 \pi T t}$ and $\Delta V/V_0 \ll 1$, we find
\begin{eqnarray}
\Delta t &\sim& \frac{\Delta V}{2 \pi T V_0} \nonumber \\
&\sim& \frac{\rho_*}{2 \pi T z_0}  = \frac{1}{4 \pi T}
\left(\frac{R_0}{z_0}\right)^{1/3}  = \frac{1}{4 (\pi T)^{2/3}}
\left( \frac{ \tilde R_0}{0.15} \right)^{1/3} \quad \textrm{for
$\rho = \bar \rho$} \,.
\end{eqnarray}
 When the tube collapses, the motion can no longer be described using
the classical equations of motion for the D7-brane.  However, one
expects that the collapsed tube breaks off from the rest of the D7-brane, perhaps annihilating in a shower of radiation.  What remains
is a D7-brane still somewhat distorted from its equilibrium
configuration, which then oscillates in $z$ until the remaining
energy is dispersed.

Let us briefly comment on larger bubbles, beyond what can be
accurately described in our Rindler approximation.  Recall that $z_0
\gtrsim 0.15 L$ so that $R_0 \gg z_0$ implies $R_0 \gg L$.  Thus,
the dynamics of such bubbles will proceed slowly enough to be
affected by the shape of the branes in the asymptotic region of the
spacetime. It is just this asymptotic region which controls the
difference in energy between the most stable black hole and
Minkowski embeddings found in \cite{MMT1,MMT2}. Since we work at a
temperature where the Minkowski embedding is thermodynamically
stable, it is clear that the energetics of this region cause the
bubble to contract; i.e., the bubble is compressed by thermodynamic
pressure.

While our Rindler approximation does not allow us to calculate the
precise timescale for this collapse, the collapse is given by the
motion of a probe D7-brane in the black D3-brane background.  As a
result, the collapse timescale must be of the form $T^{-1}$
times a function of $R_0$, the initial height $z_0$, and the AdS
scale $L$. The multiplicative factor of $T^{-1}$ comes from the
metric (\ref{rindler}) and appears when translating proper time to
coordinate time.  The proper time, however, cannot depend on the
temperature $T$ since, in the particular model considered, one may
change $T$ via a change of coordinates. In gauge theory terms, this
is associated with the scale invariance of the $N=4$
super-Yang-Mills theory to which the quarks have been added as a
small perturbation. From the gauge theory perspective the timescale
is determined by $T, T_f$, and the bubble size $\tilde R_0$ and is
otherwise independent of $\lambda, N_c,N_f$.  Because additional
dependence on $T$ appears when expressing $z_0$ in terms of gauge
theory parameters, $\Delta t$ need not be proportional to $T^{-1}$.

It is also interesting to analyze the bubble dynamics directly in
gauge theory terms.  In contrast to the above, we now assume that
the 't Hooft coupling $\lambda$ is small in order that we may
calculate reliably.

In the gauge theory, the bubble of melted mesons is distinguished
from its surrounding plasma by the presence of deconfined quarks.
 The bubble shrinks due to processes where these quarks annihilate to
form gluons which may then leave the surface of the bubble and carry
away energy.  Quarks can also combine to form mesons, but this
process is suppressed by a factor of $1/N_c$ in the large $N_c$
limit. The quark-quark-gluon vertex factor is $g_{YM} =
\sqrt{\lambda/N_c}$ so the production rate scales as
 \be
  N_f N_c^2
 \left(\sqrt{\frac{\lambda}{N_c}}\right)^2 \sim \lambda N_f N_c \,,
 \ee
in the limit of large $N_c$.  Here the factor $N_f N_c^2$ comes from
summing over initial states.  Since there are $N_f$ species of heavy
quarks, and since each species comes in $N_c$ colors, the lifetime
of the bubble scales as $N_f^0 N_c^0$.  This result matches the
scaling with $N_c,N_f$ found on the gravity side at large $\lambda$,
where collapse of the bubble was merely classical brane dynamics in
a fixed background.

One may expect the bubble to decay only due to interactions
near the surface so that the lifetime of the bubble scales with the
volume-to-surface area ratio $\tilde R_0$.  This expectation is
justified when the interior of the bubble is in local thermal
equilibrium so that it is equally likely there for two quarks to
combine into a gluon, or a gluon to decay into two quarks. However,
since it exists only for a time $\Delta t \ll 1/T$, our bubble is in
fact quite far from thermal equilibrium and it is possible that the
entire bubble contributes to the decay of the bubble.

 \section{Discussion}

The above sections used gauge/gravity duality to describe two types
of processes involving the rapid transfer of localized energy to
deconfined plasmas at large $\lambda, N_c$.  The first process
involves high energy collisions with balls of deconfined plasma
surrounded by a confining phase, while the second involves the
localized heating of a deconfined plasma.  Both processes correspond
to collisions with black holes in the gravity dual, where they
result in the nucleation of a new ``arm'' of the horizon that
reaches out in the direction of the incident object.  In the first
case, the arm stretches into a gauge theory direction while in the
second it extends in the holographic direction.   We have described
the shape of this arm in the limit where the incident object is
highly boosted and where one considers only the region close to the
original black hole. After the collision, the arm is absorbed into
the body of the black hole, though we have not described this
process in any detail.

Within the context of this duality, our results show that highly
boosted particles incident on a plasma ball induce the formation of
``virtual arms" which, though they contain no plasma themselves,
will with large probability cause any additional particles passing
through them to be captured by the plasma (see
Fig.~\ref{newplasmafig}). These virtual arms summarize the
interactions of both the original plasma ball and the incident
highly boosted particle with any additional particles that may be
present. The shape of a virtual arm was determined in section
\ref{pbs} (see equation (\ref{PBlength})), where the arm length was
found to scale with the energy of the incident particle and
inversely with a power of the cut-off $\tilde r$ on the transverse
size.

It is straightforward to generalize our single-particle analysis to
the case of a dense cluster of particles, such as might be present
in a pulsed beam.  There are two essential points: {\it i}) The arm
shape is driven by the Aichelburg-Sexl potential which is determined
by solving a Poisson equation whose source is the incident energy.
Thus, this potential satisfies a superposition principle.  {\it ii})
Since the dynamical timescale for the horizon is $1/T$, the response
of the horizon is independent of the distribution of the incident
energy along the direction of motion as long as the entire pulse of
energy arrives in a time $\Delta t \ll 1/T$.  Thus, (\ref{PBlength})
may be used to calculate the length of the virtual arm induced
collectively by particles in the cluster by taking the energy $E$ to
include all incident particles that arrive within the relaxation
timescale $1/T$ and $r$ to describe the size of the incoming cluster
of particles.  In fact, if the cluster is rotationally invariant
about the direction of motion, (\ref{PBlength}) can be used to
compute the arm shape by taking $E(r)$ to be the energy contained
within a cylinder of radius $r$.

The scaling with $E$ is universal in the limit of a thin
relativistic beam, independent of the detailed model for the
duality. The scaling with the cut-off $r$ depends only on the
dimension of the gravity dual.  It would therefore be interesting to
compare both scalings with calculations in a weakly coupled gauge
theory and with experiments that might one day be performed with QCD
plasmas. Such experiments could provide a new test of the extent to
which QCD plasmas might be described by a classical gravity dual,
independent of the details of any particular model.

We also obtained results in the context of local heating of plasmas.
Not surprisingly, we found that such heating can produce bubbles of
a higher temperature phase, but that these bubbles are unstable and
collapse.  In terms of the proper bubble size $R_0$ in the bulk or
the proper bubble size $\tilde R_0$ in the gauge theory, we were
able to study the dynamics in detail for $T \sim T_f$, large
$\lambda$, and small bubbles $R_0 \ll z_0,$ $\tilde R_0 \ll 1/T$.
There we found that the bubbles are surrounded by a much larger
region where meson propagation is highly disrupted. We also found
that the bubbles collapse within a time $\Delta t \sim \frac{1}{4
(\pi T)^{2/3}} \left( \frac{ \tilde R_0}{0.15} \right)^{1/3}.$  More
generally (for larger $\tilde R_0$, smaller $T$, or small $\lambda$)
the collapse time may depend on $T,T_f,$ and $\tilde R_0$, but must
remain otherwise independent of $\lambda, N_f,N_c$.

A result of particular interest is the dependence of the bubble
radius on both the energy $E$ and the temperature $T$.  Recall that
we considered an example four-dimensional gauge theory with $N_f$
flavors of fundamental matter in which the mesons were stable 
at $T < T_f$, but where the bubble corresponded to a phase in which
mesons are melted.  It is clear that little of the incident energy
is transferred to the mesons, so the main effect is a local heating
of the plasma.  Since the specific heat of the system is of order
$N_c^2$ and the mesons melt at temperature $T_f$, one might
therefore expect the bubble size to be of order
 \begin{equation}
\tilde R_{naive} \sim \left( \frac{E } { (T_f -T)T_f^3 N_c^2}
\right)^{1/3}
 \end{equation}
as there is only enough energy to raise the temperature from $T$ to
$T_f$ over a region of this size.

However, at least when $T \sim T_f$, the bubbles that form in our
processes are in fact of size $\tilde R_0 \sim \frac{E} {\pi (0.15)^2 T_f^2
N_c^2}$, where we have translated (\ref{R0}) to gauge theory
parameters using $\tilde R_0 = R_0/(\pi T L)$, $z_0 \sim L$ and
(\ref{L4}). Thus, the bubble size is independent of $T-T_f$ and, in
the limit of large $E$, the actual bubble size $\tilde R_0$ is much
{\it larger} than $\tilde R_{naive}$. This is due to the fact that
the bubble is quite far from thermal equilibrium. In fact, from the
bulk perspective the black hole arm that pulls the probe brane
behind the horizon exists only for the time scale $\Delta t \sim
1/T$ typical of thermal fluctuations.  Thus, there is no reason to
assign the arm an energy density characteristic of thermal
equilibrium at temperature $T_f$.

Our analysis considered only physics associated with the region
close to the original black hole horizon and cases where the
incident particles approach the black hole at relativistic speeds.
In such regimes, the behavior of the black hole horizon is universal
and independent of the details of any particular duality.  One may
therefore hope that this limit describes universal physics of
deconfined plasmas at large $N_c$.  However, it is also clear that
there is much to learn from the details of particular models in
regimes where these limits do not apply.  Some such extensions of
this work may be straightforward, such as taking into account the
details of the black D3-brane solution and the various D7-brane
embeddings found in \cite{MMT1,MMT2}.  Other more complicated
extensions might involve studying the collisions of two plasma
balls.  Such situations are also more complicated  from the bulk
viewpoint and may require sophisticated numerical simulations.
However, recent progress (see e.g., \cite{FP}) in simulating black
hole collisions in 3+1 dimensions suggests that such calculations
may nevertheless prove tractable in the near future.

 \label{disc}

 \subsection*{Acknowledgements}
 We thank David Mateos for his patient explanations of AdS/QCD with
flavor and for many enlightening discussions. AV would also like to
thank Anshuman Maharana and Matthew Roberts for several useful
discussions. This work was supported in part by the National Science
Foundation under Grant No.~PHY05-55669, and by funds from the
University of California.

\appendix

\section{Brane dynamics in the Aichelburg-Sexl solution}
 \label{app}
In this appendix we provide a brief calculation of the D7-brane's
initial velocity induced by crossing the Aichelburg-Sexl shock wave.
We find that there is a region between  $z = 0$ and $z \sim z_0 -
(R_0^2 z_0)^{1/3}$  where the bubble nucleates in a contracting
phase, as claimed in section \ref{melt}

In analogy to the solutions for null geodesics found in section
\ref{setup}, let us assume that the D7-brane configuration is
described by specifying $U(V,\rho)$. Hence, the induced metric on
the brane takes the form
\begin{equation}
ds^2_{brane} = \left(-\frac{\partial U}{\partial V} + \mu \Phi(\rho)
\delta(V-V_0) \right) dV^2+ \frac{\partial U}{\partial \rho} d\rho
dV + d\rho^2+ \rho^2 d\Omega_2^2+dy^2+y^2 d\Omega_3^2 \,.
\end{equation}
As a result, the Lagrangian density appearing in the DBI
action for the brane is
\begin{equation}
\label{dbi} \mathcal{L} \propto \rho^2 \sqrt{\frac{\partial
U}{\partial V}+\frac{1}{4} \left(\frac{\partial U}{\partial
\rho}\right)^2- \mu \Phi(\rho) \delta(V-V_0) } \,.
\end{equation}
Based on the discussion in section \ref{melt}, we choose an ansatz
for $U(V,\rho)$ of the form
\begin{equation}
U(V, \rho) = -\frac{z_0^2}{V}+ h(V,\rho) \Theta(V-V_0) \,,
\end{equation}
where $h(V,\rho)$ is continuous in $V$ and satisfies $h(V_0,\rho) =
|\Delta U| = \mu \Phi(\rho)$. Using this ansatz and the equation of
motion obtained from (\ref{dbi}), we find that immediately after the shock
wave passes 
\begin{equation}
\frac{\partial U}{\partial V} = \frac{z_0^2}{V_0^2}
+\left.\frac{\partial h}{\partial V}\right|_{V = V_0} =
\frac{z_0^2}{V_0^2} -\frac{\mu^2 \Phi'(\rho)^2}{8} \,.
\end{equation}
Now, recall that the Rindler coordinate $z$ is related to $U,V$
through $z = \sqrt{-U V}$, so the velocity in the $z$ direction (at
fixed $\rho$) is
\begin{equation}
\left(\frac{\partial z}{\partial V}\right)_\rho = -\frac{1}{2
\sqrt{-U V}} \left(U + V \left(\frac{\partial U}{\partial
V}\right)_\rho \right) \,.
\end{equation}
The radial velocity of the cylinder (at fixed $z$) is then given by
\begin{equation}
\label{rdot}
 \left(\frac{\partial \rho}{\partial V}\right)_z =
-\left(\frac{\partial z}{\partial V}\right)_\rho \,
\left(\frac{\partial z}{\partial \rho}\right)_V^{-1}
 = \frac{1}{V_0} \left(\rho - \frac{R_0 z_0^2}{8\rho^2} \right)
 \, ,
\end{equation}
where we have used $z(V_0, \rho) = \ z_0 \sqrt{1-R_0/\rho}$ and
$R_0$ is defined in (\ref{R0}).  Equation (\ref{rdot}) implies that
$\partial \rho /\partial V$ is indeed negative for $\rho <
\rho_* \equiv \frac{(R_0 z_0^2)^{1/3}}{2}$, though it is positive
for $\rho
> \rho_*$.


\begin{thebibliography}{99}

 %%%%%%%%%%%%%%%%%%%%%%%%%%%%%%%%%% ADS/CFT %%%%%%%%%%%%%%%%%%%%%%%%%%%%%%%

\bibitem{Juan}
J.~M.~Maldacena, ``The large N limit of superconformal field
theories and supergravity," Adv.~Theor.~Math.~Phys. {\bf 2} (1998)
231 [Int. J. Theor. Phys. 38 (1999) 1113] [arXiv:hep-th/9711200].

\bibitem{MAGOO}
O.~Aharony, S.~S.~Gubser, J.~M.~Maldacena, H.~Ooguri, and Y.~Oz,
``Large N field theories, string theory and gravity," Phys.~Rept.
{\bf 323} (2000) 183 [arXiv:hep-th/9905111].

\bibitem{wit}
E.~Witten, ``Anti-de Sitter space, thermal phase transition, and
confinement in gauge theories," Adv.~Theor.~Math.~Phys. {\bf 2}
(1998) 505 [arXiv: hep-th/9803131].


%%%%%%%%%%%%%%%%%%%%% SHEAR VISCOSITY AND SOUND WAVES %%%%%%%%%%%%%%%%%%%%%%%%%%%%%%%

\bibitem{shear_viscosity}

G.~Policastro, D.~T.~Son, and A.~O.~Starinets, ``The shear viscosity
of strongly coupled N = 4 supersymmetric Yang-Mills plasma,''
Phys.\ Rev.\ Lett.\  {\bf 87}, 081601 (2001)  [arXiv:hep-th/0104066]. \\

G.~Policastro, D.~T.~Son, and A.~O.~Starinets,``From AdS/CFT
correspondence to hydrodynamics,'' JHEP {\bf 0209}, 043 (2002)
[arXiv:hep-th/0205052]. \\

G.~Policastro, D.~T.~Son, and A.~O.~Starinets,``From AdS/CFT
correspondence to hydrodynamics II: Sound waves,'' JHEP {\bf 0212},
054 (2002)
[arXiv:hep-th/0210220]. \\

C.~P.~Herzog,
``The hydrodynamics of M-theory,''  JHEP {\bf 0212}, 026 (2002)  [arXiv:hep-th/0210126]. \\

C.~P.~Herzog,
``The sound of M-theory,'' Phys.~Rev.~D {\bf 68} (2003) 024013 [arXiv:hep-th/0302086]. \\

A.~Buchel and J.~T.~Liu, ``Universality of the shear viscosity in
supergravity,'' Phys.\ Rev.\ Lett.\  {\bf 93}, 090602 (2004)
[arXiv:hep-th/0311175]. \\

P.~Kovtun, D.~T.~Son, and A.~O.~Starinets, ``Holography and
hydrodynamics: Diffusion on stretched horizons,''  JHEP {\bf 0310},
064 (2003)
[arXiv:hep-th/0309213]. \\

P.~Kovtun, D.~T.~Son, and A.~O.~Starinets, ``Viscosity in strongly
interacting quantum field theories from
black hole  physics,''  Phys.\ Rev.\ Lett.\  {\bf 94}, 111601 (2005) [arXiv:hep-th/0405231]. \\

A.~Buchel, ``On universality of stress-energy tensor correlation
functions in supergravity,''
Phys.\ Lett.\  B {\bf 609}, 392 (2005)  [arXiv:hep-th/0408095]. \\

P.~Benincasa, A.~Buchel, and A.~O.~Starinets, ``Sound waves in
strongly coupled non-conformal gauge theory plasma,''  Nucl.\ Phys.\
B {\bf 733}, 160 (2006)  [arXiv:hep-th/0507026]. \\

For further references see \cite{MatRev}.

%%%%%%%%%%%%%%%%%%%%%%%%%%%%%%%%%%% BULK VISCOSITY  %%%%%%%%%%%%%%%%%%%%%%%%%%%%%%%%%%

\bibitem{bulk_viscosity}
A.~Parnachev and A.~Starinets, ``The silence of the little
strings,''  JHEP {\bf 0510}, 027 (2005)
[arXiv:hep-th/0506144]. \\

P.~Benincasa, A.~Buchel, and A.~O.~Starinets, ``Sound waves in
strongly coupled non-conformal gauge theory plasma,''  Nucl.\ Phys.\
B {\bf 733}, 160 (2006)
[arXiv:hep-th/0507026]. \\

J.~Mas and J.~Tarrio,
``Hydrodynamics from the Dp-brane,''  JHEP {\bf 0705}, 036 (2007) [arXiv:hep-th/0703093]. \\

A.~Buchel, ``Bulk viscosity of gauge theory plasma at strong
coupling,'' arXiv:0708.3459 [hep-th] and references therein.

%%%%%%%%%%%%%%%%%%%%%%%%%%%%%%%%% JET QUENCHING %%%%%%%%%%%%%%%%%%%%%%%%%

\bibitem{jet}
H.~Liu, K.~Rajagopal, and U.~A.~Wiedemann, ``Calculating the jet
quenching parameter from AdS/CFT,''  Phys.\ Rev.\ Lett.\  {\bf 97},
182301 (2006)
[arXiv:hep-ph/0605178].  \\

C.~P.~Herzog, A.~Karch, P.~Kovtun, C.~Kozcaz, and L.~G.~Yaffe,
``Energy loss of a heavy quark moving through N = 4 supersymmetric
Yang-Mills plasma,''  JHEP {\bf 0607},
013 (2006)  [arXiv:hep-th/0605158]. \\

J.~Casalderrey-Solana and D.~Teaney,
``Heavy quark diffusion in strongly coupled N = 4 Yang-Mills," Phys.~Rev.~D {\bf 74} 085012 (2006) [arXiv:hep-ph/0605199]. \\

S.~S.~Gubser, ``Drag force in AdS/CFT," Phys.~Rev.~D {\bf 74} 126005 (2006) [arXiv:hep-th/0605182]. \\

For a review see:  H.~Liu,``Heavy ion collisions and AdS/CFT,''  J.\
Phys.\ G {\bf 34}, S361 (2007) [arXiv:hep-ph/0702210].

%%%%%%%%%%%%%%%%%%%%%%%%%%%%%%%%%%%%%%% FAR FROM EQUILIBRUIM %%%%%%%%%%%%%%%%%%%%%%%%%%%



\bibitem{far}

See for example:\\

H.~Nastase,
``The RHIC fireball as a dual black hole,'' [arXiv:hep-th/0501068]. \\

E.~Shuryak, S.~J.~Sin, and I.~Zahed, ``A gravity dual of RHIC
collisions,''  J.\ Korean Phys.\ Soc.\  {\bf
50}, 384 (2007)  [arXiv:hep-th/0511199]. \\

R.~A.~Janik and R.~Peschanski, ``Asymptotic perfect fluid dynamics
as a consequence of AdS/CFT,''  Phys.\ Rev.\  D {\bf 73}, 045013
(2006)
[arXiv:hep-th/0512162]. \\


H.~Nastase, ``More on the RHIC fireball and dual black holes,''
[arXiv:hep-th/
0603176]. \\


R.~A.~Janik and R.~Peschanski, ``Gauge / gravity duality and
thermalization of a boost-invariant perfect fluid,''  Phys.\ Rev.\ D
{\bf 74}, 046007 (2006)
[arXiv:hep-th/0606149].  \\



R.~A.~Janik, ``Viscous plasma evolution from gravity using
AdS/CFT,'' Phys.\ Rev.\ Lett.\  {\bf 98}, 022302
(2007)  [arXiv:hep-th/0610144]. \\

S.~Lin and E.~Shuryak, ``Toward the AdS/CFT gravity dual for high
energy heavy ion collisions,''
[arXiv:hep-ph/0610168]. \\

J.~J.~Friess, S.~S.~Gubser, G.~Michalogiorgakis, and S.~S.~Pufu,
``Expanding plasmas and quasinormal modes of anti-de Sitter black
holes,''  JHEP {\bf 0704}, 080 (2007)
[arXiv:hep-th/0611005]. \\


E.~Shuryak, ``Emerging theory of strongly coupled quark-gluon
plasma,''  [arXiv:hep-ph/0703208]. \\

M.~P.~Heller and R.~A.~Janik,
 ``Viscous hydrodynamics relaxation time  from AdS/CFT,''  Phys.\
Rev.\  D {\bf 76}, 025027 (2007)  [arXiv:hep-th/0703243].

%%%%%%%%%%%%%%%%%%%%%%%%%%%%%%%%% MATEOS REVIEW %%%%%%%%%%%%%%%%%%%%%%%%%%%%%%

\bibitem{MatRev}
D.~Mateos, ``String Theory and Quantum Chromodynamics,''
arXiv:0709.1523 [hep-th].

%%%%%%%%%%%%%%%%%%%%%%%%%%%%%%%%%%%% EQUILIBRIUM %%%%%%%%%%%%%%%%%%%%%%%%%%

\bibitem{equilibrium}
S.~S.~Gubser, I.~R.~Klebanov, and A.~W.~Peet,``Entropy and
Temperature of Black 3-Branes,''  Phys.\ Rev.\  D {\bf 54}, 3915
(1996) [arXiv:hep-th/9602135]. \\
S.~S.~Gubser, I.~R.~Klebanov, and A.~A.~Tseytlin, ``Coupling
constant dependence in the thermodynamics of N = 4  supersymmetric
Yang-Mills theory,''  Nucl.\ Phys.\  B {\bf 534}, 202 (1998)
[arXiv:hep-th/9805156].


%%%%%%%%%%%%%%%%%%%%%%%% ELLIPTIC FLOW %%%%%%%%%%%%%%%%%%%%%%%%%

\bibitem{Shuryak}

D.~Teaney,
``Effect of shear viscosity on spectra, elliptic flow, and Hanbury Brown-Twiss radii," Phys.~Rev.~C {\bf 68} 034913 (2003) [arXiv:nucl-th/0301099]. \\

E.~Shuryak, ``Why does the quark gluon plasma at RHIC behave as a
nearly ideal fluid?,''  Prog.\ Part.\ Nucl.\ Phys.\  {\bf 53}, 273
(2004)  [arXiv:hep-ph/0312227].



\bibitem{membrane}
``Black Holes: The Membrane Paradigm," R. Price, K. Thorne,
D.A.MacDonald, eds. (Yale University Press, 1986).

\bibitem{US}
U. Sperhake et al, in preparation.

\bibitem{EarlyNumerics}
S.~L.~Shapiro and S.~A.~Teukolsky,
  ``Collisions of relativistic clusters and the formation of black holes,''
  Phys.\ Rev.\  D {\bf 45}, 2739 (1992). \\
  %%CITATION = PHRVA,D45,2739;%%
 P.~Anninos, D.~Hobill, E.~Seidel, L.~Smarr, and W.~M.~Suen,
  ``The collision of two black holes,''
  Phys.\ Rev.\ Lett.\  {\bf 71}, 2851 (1993)
  [arXiv:gr-qc/9309016]. \\
  %%CITATION = PRLTA,71,2851;%%
 S.~A.~Hughes, C.~R.~Keeton, P.~Walker, K.~T.~Walsh, S.~L.~Shapiro, and S.~A.~Teukolsky,
  ``Finding black holes in numerical space-times,''
  Phys.\ Rev.\  D {\bf 49}, 4004 (1994). \\
  %%CITATION = PHRVA,D49,4004;%%
  A.~M.~Abrahams, G.~B.~Cook, S.~L.~Shapiro, and S.~A.~Teukolsky,
  ``Solving Einstein's equations for rotating space-times: Evolution of
  relativistic star clusters,''
  Phys.\ Rev.\  D {\bf 49}, 5153 (1994). \\
  %%CITATION = PHRVA,D49,5153;%%
  P.~Anninos {\it et al.},
  ``Dynamics Of Apparent And Event Horizons,''
  Phys.\ Rev.\ Lett.\  {\bf 74}, 630 (1995)
  [arXiv:gr-qc/9403011]. \\
  %%CITATION = PRLTA,74,630;%%
 J.~Libson, J.~Masso, E.~Seidel, W.~M.~Suen, and P.~Walker,
  ``Event horizons in numerical relativity. 1: Methods and tests,''
  Phys.\ Rev.\  D {\bf 53}, 4335 (1996)
  [arXiv:gr-qc/9412068]. \\
  %%CITATION = PHRVA,D53,4335;%%
   R.~A.~Matzner, H.~E.~Seidel, S.~L.~Shapiro, L.~Smarr, W.~M.~Suen, S.~A.~Teukolsky, and J.~Winicour,
  ``Geometry of a black hole collision,''
  Science {\bf 270}, 941 (1995). \\
  %%CITATION = SCIEA,270,941;%%
  S.~L.~Shapiro, S.~A.~Teukolsky, and J.~Winicour,
  ``Toroidal Black Holes And Topological Censorship,''
  Phys.\ Rev.\  D {\bf 52}, 6982 (1995).
  %%CITATION = PHRVA,D52,6982;%%


% \bibitem{FP} F.~Pretorius,``Binary Black Hole Coalescence,''  arXiv:0710.1338 [gr-qc].  %%CITATION = ARXIV:0710.1338;%%

\bibitem{HE} S.~W.~Hawking and G.~F.~R.~Ellis, ``The large scale structure of space-time,'' Cambridge University Press (1973).

\bibitem{Waldbook}
R.~M.~Wald, ``General Relativity," University of Chicago Press
(1984).


\bibitem{AMV}
A.~J.~Amsel, D.~Marolf, and A.~Virmani,
``The Physical Process First Law for Bifurcate Killing Horizons,'' arXiv:0708.2738 [gr-qc].%%CITATION = ARXIV:0708.2738;

\bibitem{AS}
P.~C.~Aichelburg and R.~U.~Sexl, ``On the Gravitational Field of a
Massless Particle,'' Gen.~Rel.~Grav., {\bf 2}, 303-312 (1971).

\bibitem{FPV}
 V.~Ferrari, P.~Pendenza, and G.~Veneziano,
 ``Beam-Like Gravitational Waves and Their Geodesics,'' Gen.~Rel.~Grav., {\bf 20}, 1185-1191 (1988).

\bibitem{DtH}
T.~Dray and G.~'t Hooft, ``The Gravitational Shock Wave of a
Massless Particle," Nucl.~Phys.~B, {\bf 253}, 173-188 (1985).

\bibitem{AMW}
O.~Aharony, S.~Minwalla, and T.~Wiseman, ``Plasma-balls in large N
gauge theories and localized black holes,''  Class.\ Quant.\ Grav.\
{\bf 23}, 2171 (2006)  [arXiv:hep-th/0507219].
 %%CITATION = CQGRD,23,2171;%%


\bibitem{HMsol}
G.~T.~Horowitz and R.~C.~Myers,
  ``The AdS/CFT correspondence and a new positive energy conjecture for
  general relativity,''
  Phys.\ Rev.\  D {\bf 59}, 026005 (1999)
  [arXiv:hep-th/9808079].
  %%CITATION = PHRVA,D59,026005;%%

\bibitem{KR}
A.~Karch and L.~Randall, ``Open and closed string interpretation of
SUSY CFT's on branes with boundaries,'' JHEP {\bf 0106} (2001) 063
[arXiv: hep-th/0105132].

\bibitem{KK}
A.~Karch and E.~Katz, ``Adding flavor to AdS/CFT,'' JHEP {\bf 0206}
(2002) 043 [arXiv: hep-th/0205236].


%%%%%%%%%%%%%%%%%%%%%%%%%% MMT Papers %%%%%%%%%%%%%%%%%%%%%%%%%%%%%%%%

\bibitem{MMT1}
 D.~Mateos, R.~C.~Myers, and R.~M.~Thomson,
``Holographic phase transitions with fundamental matter,"
Phys.~Rev.~Lett. {\bf 97} (2006) 091601 [arXiv: hep-th/0605046].

\bibitem{MMT2}
D.~Mateos, R.~C.~Myers, and R.~M.~Thomson, ``Thermodynamics of the
brane,"  JHEP {\bf 0705}, 067 (2007) [arXiv: hep-th/0701132].



\bibitem{fro}
V.~P.~Frolov, ``Merger transitions in brane-black-hole systems:
Criticality, scaling, and self-similarity," Phys.~Rev.~D  {\bf 74}
(2006) [arXiv: gr-qc/0604114].

\bibitem{GL}
 R.~Gregory and R.~Laflamme,
  ``Black strings and p-branes are unstable,''
  Phys.\ Rev.\ Lett.\  {\bf 70}, 2837 (1993)
  [arXiv:hep-th/9301052].
  %%CITATION = PRLTA,70,2837;%%


\bibitem{FP}
F.~Pretorius,
  ``Binary Black Hole Coalescence,''
  arXiv:0710.1338 [gr-qc].
  %%CITATION = ARXIV:0710.1338;%%

 \end{thebibliography}
 \end{document}